# Direct observation of vortices in an electron fluid

A. Aharon-Steinberg[1,†], T. Völkl[1,†], A. Kaplan[1,†], A. K. Pariari[1], I. Roy[1], T. Holder[1], Y. Wolf[1], A. Y. Meltzer[1], Y. Myasoedov[1], M. E. Huber[2], B. Yan[1], G. Falkovich[3], L. S. Levitov[4], M. Hücker[1], and E. Zeldov[1,*]

Vortices are the hallmarks of hydrodynamic flow. Recent studies indicate that strongly-interacting electrons in ultrapure conductors can display signatures of hydrodynamic behavior including negative nonlocal resistance [1–6], Poiseuille flow in narrow channels [7–10], and a violation of the Wiedemann-Franz law [11,12]. Here we provide the first visualization of whirlpools in an electron fluid. By utilizing a nanoscale scanning superconducting quantum interference device on a tip (SQUID-on-tip) [13] we image the current distribution in a circular chamber connected through a small aperture to an adjacent narrow current-carrying strip in high-purity type-II Weyl semimetal $WTe_2$. In this geometry, the Gurzhi momentum diffusion length and the size of the aperture determine the vortex stability phase diagram. We find that the vortices are present only for small apertures, whereas the flow is laminar (non-vortical) for larger apertures, consistent with the theoretical analysis of the hydrodynamic regime and in contrast to the expectations of ballistic transport in $WTe_2$ at low temperatures [10]. Moreover, near the vortical-to-laminar transition, we observe a single vortex in the chamber splitting into two vortices, a behavior that can occur only in the hydrodynamic regime and cannot be sustained by ballistic transport. These findings suggest a novel mechanism of hydrodynamic flow: instead of the commonly considered electron-electron scattering in the bulk, which becomes extremely weak at low temperatures, the spatial diffusion of the charge carriers' momenta is enabled by small-angle scattering at the planar surfaces of thin pure crystals. This surface-induced para-hydrodynamics opens new avenues for exploring and utilizing electron fluidics in high-mobility electron systems.

_________________________________

[1]Department of Condensed Matter Physics, Weizmann Institute of Science, Rehovot 7610001, Israel

[2]Departments of Physics and Electrical Engineering, University of Colorado Denver, Denver, Colorado 80217, USA

[3]Department of Physics of Complex Systems, Weizmann Institute of Science, Rehovot 7610001, Israel

[4]Department of Physics, Massachusetts Institute of Technology, Cambridge, Massachusetts 02139, USA

[†]These authors contributed equally to this work

[*]eli.zeldov@weizmann.ac.il

Recent years have seen a quest for systems and regimes in which strong electron-electron interactions may lead to electron flows governed by hydrodynamics [14], as in viscous fluids, rather than by Ohmic transport. Fluids display two distinct hydrodynamic regimes [15]: laminar flows in which neighboring sheets move at gradually varying velocities, and turbulent flows characterized by eddies and vortices with counterflow that develop into strongly fluctuating and chaotic behavior at large scales. The transition from laminar to turbulent flow is usually associated with nonlinear fluid dynamics, described by the Navier–Stokes equations. Yet, already in linear Stokes flow, hydrodynamic vortices in Newtonian fluids readily occur [16].

In contrast to common fluids, which display hydrodynamic phenomena abundantly, evidence for hydrodynamics in electron fluids has remained scarce [17,18] until recently. The advent of high purity single crystals, clean van der Waals heterostructures, and high mobility 2D systems has accelerated the observation of fluid-like behavior in semiconductors and semimetals [1–12,19–29] and triggered a flurry of theoretical works [19,30–61]. Recently, the laminar Poiseuille flow in narrow strips has been demonstrated in graphene [7–9] and in WTe$_2$ [10] with the help of Hall potential imaging and diamond nitrogen-vacancy magnetometry, lending support to the hydrodynamic nature of electron fluids in these systems. Yet, the most striking and ubiquitous feature in the flow of regular fluids, the formation of vortices and turbulence, has not yet been observed in electron fluids despite numerous theoretical predictions based on linear [62–64] and nonlinear [33,39,65] hydrodynamics. Transport measurements showing negative nonlocal resistance in the vicinity of the current injection point are suggestive of electron backflow in graphene and GaAs heterostructures [1–6]. Recent studies, however, propose that the observed negative potentials may arise in ballistic and hydrodynamic regimes even without an actual electron backflow [3,36,40]. Hence, direct observation of vortices and the study of their properties remains an outstanding challenge in electron fluids.

The conventional picture of electron transport [14] involves two distinct length scales -- the momentum-relaxation length $l_{mr}$ describing momentum transfer from electrons to the lattice and the length $l_{ee}$ describing momentum transfer between the carriers due to electron-electron collisions. Hydrodynamic behavior can harbor vortices and other unique effects associated with electron fluidity when $l_{ee} \ll l_{mr}$, at the length scales $l_{ee} < W < l_{mr}$, where $W$ is the characteristic size of the system. In common metals, to the contrary, the shortest length scale is $l_{mr}$. In this case, the electron transport is described by Ohm's law $\boldsymbol{J} = -\sigma\nabla\phi = \sigma\boldsymbol{E}$ and the continuity equation $\nabla \cdot \boldsymbol{J} = 0$, where $\phi$ is the electrostatic potential and $\boldsymbol{E}$ is the electric field. In this regime, therefore, no vortices can exist since $\nabla \times \boldsymbol{J} = \sigma\nabla \times \boldsymbol{E} = 0$ in the steady state. Note that since we discuss quasi-2D geometry, $\boldsymbol{J}$ and $\sigma$ denote here the 2D current density and conductivity, respectively. In ultraclean systems at low temperatures, however, very large values of $l_{mr}$, exceeding $W$, can be achieved. If both $l_{ee}$ and $l_{mr}$ exceed $W$, the transport is ballistic, in which case electrons propagate essentially unimpeded, with scattering occurring mainly at the device edges. Ballistic transport, unlike the ohmic transport, can also lead to vortices with interesting properties.

In this work, we provide a direct visualization of vortices in an electron fluid. By utilizing magnetic imaging with a scanning superconducting quantum interference device fabricated on the apex of a sharp pipette (SQUID-on-tip, SOT) [13], we observe the current flow patterns in the Weyl semimetal WTe$_2$ as shown schematically in Figs. 1a,b. Ultraclean single crystals with residual resistance ratio ($RRR$) of over 3,000 were grown as described in Methods (Extended Data Fig. 1) and exfoliated into thin flakes of thickness $d$ of 23 to 48 nm. Various sample geometries were patterned using e-beam lithography and plasma etching (Methods). The primary geometry consists of a central strip of width $W = 550$ nm with two truncated circular chambers of radius $R = 900$ nm connected to its sides through apertures defined by the opening angle $\theta \leq 180°$(see AFM images in Fig. 1c and Extended Data Fig. 6). Analogous geometries were patterned in Au films of similar thicknesses for comparison. An *ac* current with rms amplitude $I_0$ of 1 to 400 µA was applied to the samples at $T = 4.5$ K. The corresponding out-of-plane component of the Oersted field $B_z(x, y)$ was measured by the SOT, scanning at a height $h = 50$ nm above the sample surface (see Figs. 1a,b, Methods, and Extended Data Fig. 2), where the $x$ and $y$ directions are defined in Fig. 1e.

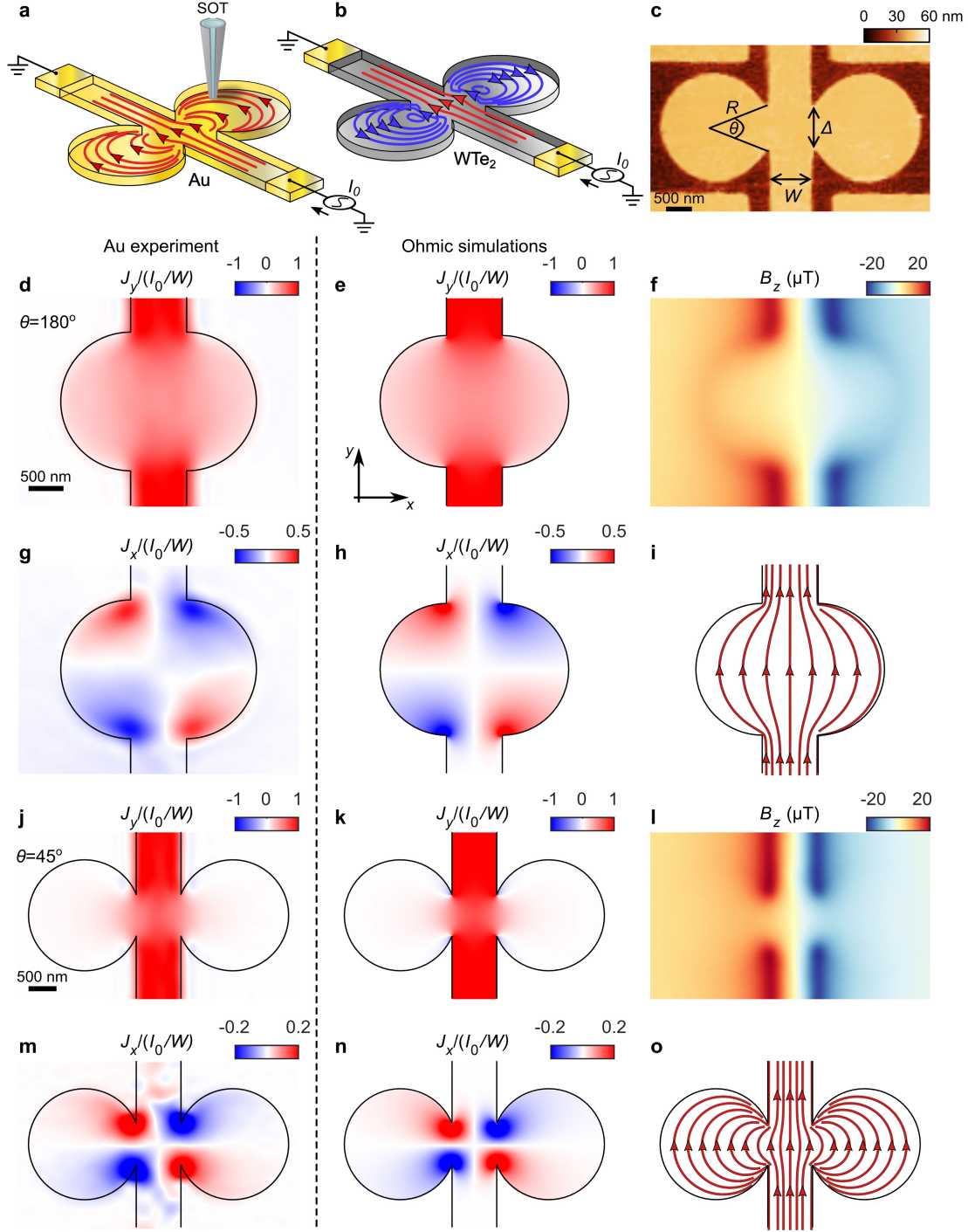

**Fig. 1. Ohmic electron flow in Au film. a-b**, Schematic experimental layout showing the scanning SOT and the Au (**a**) and WTe$_2$ (**b**) samples with double-chamber geometry. The red curves indicate laminar (open-loop) current streamlines while the blue curves represent closed-loop vortex streamlines. **c**, Atomic force microscope (AFM) topography image of WTe$_2$ sample with $\theta = 45°$. $W$, $R$, $\theta$, and $\Delta$ are the width of the central strip, radius of the circular chambers, aperture angle, and size of the aperture $\Delta = 2R\sin(\theta/2)$, respectively. **d-i**, Measurements and simulations of Au sample with $\theta = 180°$. **d**, Current density $J_y(x,y)$ normalized by $I_0/W$ reconstructed from $B_z(x,y)$ in (**f**). The black contours mark the sample edges. **e**, Simulated $J_y(x,y)$ in the ohmic regime. **f**, $B_z(x,y)$ measured by the SOT above the Au sample under current $I_0 = 50$ μA at 4.5 K. **g**, $J_x(x,y)$ reconstructed from (**f**). The light blue texture outside the sample is an artifact of current reconstruction (Methods). **h**, Simulated $J_x(x,y)$. **i**, Simulated current streamlines. **j-o**, Measurements and simulations of Au sample with $\theta = 45°$. **j**, Current density $J_y(x,y)$ reconstructed from (**l**). **k**, Simulated $J_y(x,y)$ in the ohmic regime. **l**, $B_z(x,y)$ profiles in the Au sample carrying $I_0 = 50$ μA. **m**, $J_x(x,y)$ reconstructed from (**l**). **n**, Simulated $J_x(x,y)$. **o**, Simulated current streamlines.

**Ohmic flow**

We start by examining the current flow in the Au films. Figure 1f shows $B_z(x,y)$ measured above the Au sample with $\theta = 180°$, corresponding to a strip with two half-disc chambers. By inversion of the magnetic field $B_z(x,y)$ [66], we reconstruct the 2D current density $\boldsymbol{J}(x,y)$ (Methods), with $J_x(x,y)$ and $J_y(x,y)$ components presented in Figs. 1d,g. The longitudinal $J_y$ component demonstrates that the current flowing upwards in the central strip spreads out into the two chambers. The current flows into the right chamber through its lower half (red $J_x$ in Fig. 1g), circulates counterclockwise, and exits through the upper half (blue $J_x$ in Fig. 1g). The left chamber shows a mirrored flow pattern, as expected. The COMSOL numerical simulations in the ohmic regime in Figs. 1e,h show good agreement with the experimental data, describing a laminar (non-vortical) current flow as illustrated by the calculated streamlines in Fig. 1i. Upon decreasing the aperture size, less current enters the chambers as seen by $J_y(x,y)$ in Fig. 1j for the case of $\theta = 45°$. The $J_x(x,y)$ in Fig. 1m shows qualitatively similar behavior, with current flowing counterclockwise in the right chamber. Numerical simulations of $J_y$ and $J_x$ in Figs. 1k,n agree with the experimental data; the streamlines in Figs. 1o show laminar flow, as expected.

As a quick note on terminology – usually 'laminar' means no turbulence. Here, in a linear response regime, turbulence is not encountered. And yet, vortices may or may not appear depending on the dynamical phase of the electron fluid. Indeed, vortex is a flow in which the streamlines form closed loops. In contrast, streamlines that go from source to drain without forming closed loops will be referred to hereafter as laminar.

**Vortex flow**

For large opening angle, $\theta = 120°$, the current flow pattern in a WTe$_2$ sample (Fig. 2a) looks similar to that of Au with the $J_y(x,y)$ component spreading substantially into the chambers. The corresponding $J_x(x,y)$ in Fig. 2d shows counter-clockwise flow in the right chamber and clockwise flow in the left chamber, similar to the laminar flow in Au in Figs. 1g,m. To quantify the expected behavior in the hydrodynamic regime, $l_{ee} < W < l_{mr}$, we numerically solve the linearized Navier-Stokes equation for a 2D electron fluid [1,34,62],

$$-D^2 \nabla^2 \boldsymbol{J} + \boldsymbol{J} = -\sigma \nabla \phi, \qquad (1)$$

where $D$ is the Gurzhi length, usually defined as $D = \sqrt{l_{ee} l_{mr}}/2$. The resulting $J_y(x,y)$ and $J_x(x,y)$ in Figs. 2b,e show good agreement with the experimental data. The corresponding calculated current streamlines in Fig. 2f show a laminar flow resembling the ohmic regime in Figs. 1i,o.

The flow pattern changes drastically as the aperture size becomes smaller, as illustrated in Figs. 2g,j for the case of $\theta = 20°$. The $J_y(x,y)$ remains focused in the central strip, while $J_y$ in the chambers is relatively small (Fig. 2g). An essential difference between the flow in WTe$_2$ and the ohmic flow in Au is revealed, however, upon inspecting the transversal component $J_x(x,y)$. On approaching the aperture from below, $J_x$ is initially directed to the right (red) towards the right chamber. Rather than maintaining its flow into the chamber as in Fig. 2d, $J_x$ switches its direction and flows to the left (blue). Similarly, in the top half of the chamber, $J_x$ flows out of the chamber near the aperture (blue), but into the chamber further away from the aperture (red). In other words, near the aperture the current flows counterclockwise while in the interior of the chamber, the current circulates clockwise. Therefore, a clockwise current vortex is formed in the right chamber. Simultaneously, a mirror-symmetric counterclockwise vortex appears in the left chamber. The hydrodynamic simulations of $J_y(x,y)$ and $J_x(x,y)$ in Figs. 2h,k confirm this picture, with the vortices in the two chambers represented by closed-loop streamlines (blue) in Fig. 2l (see also Extended Data Fig. 4). To the best of our knowledge, this constitutes the first direct observation of current vortices in an electron fluid.

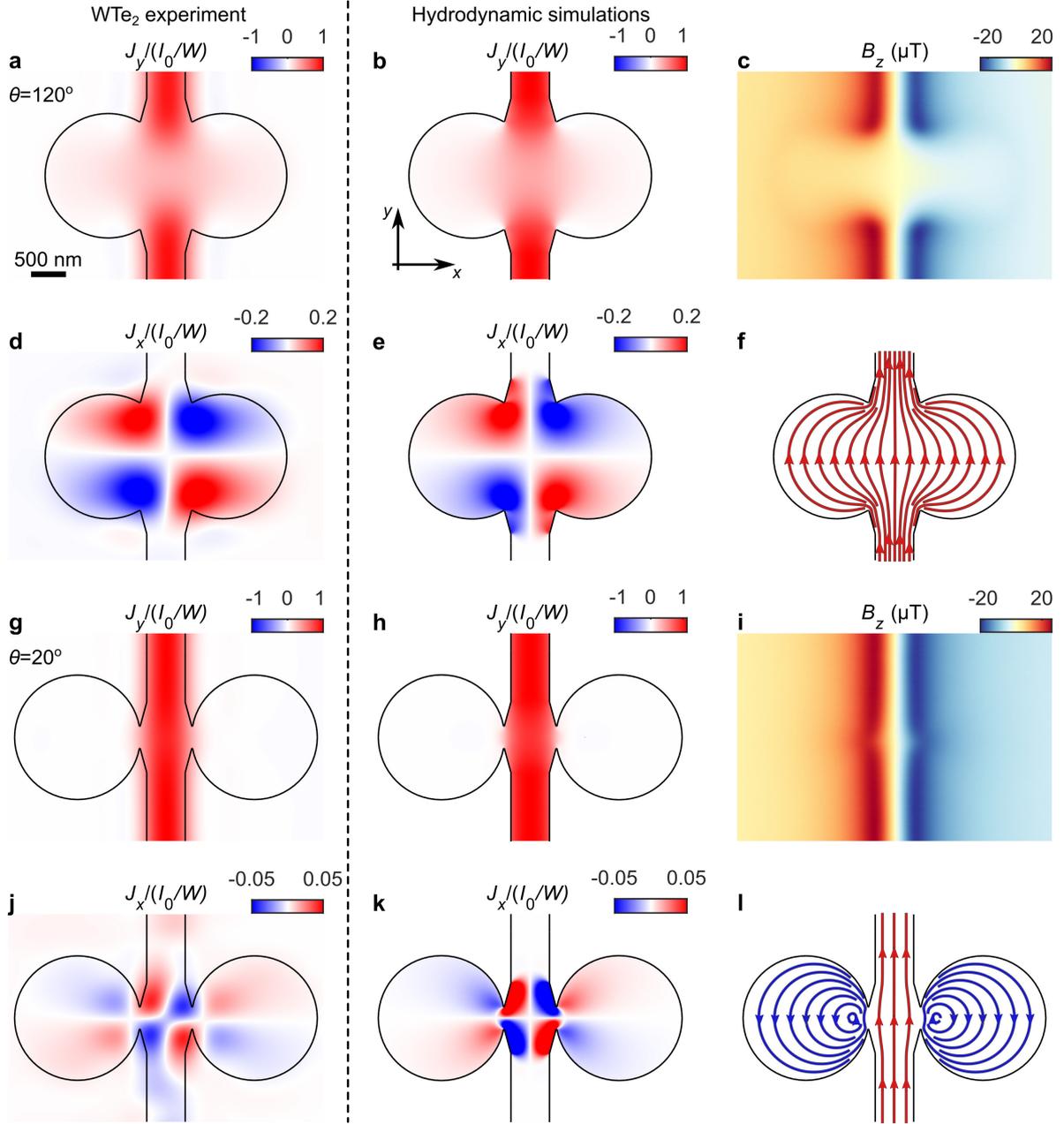

**Fig. 2**. **Laminar and vortex flow in WTe$_2$. a-f,** Measurements of WTe$_2$ sample with $\theta = 120°$ and corresponding simulations in the hydrodynamic regime (Eq. 1) with $D/W = 0.28$ and $\xi = 200$ nm. **a**, Current density $J_y(x,y)$ normalized by $I_0/W$ reconstructed from (**c**). **b**, Simulated $J_y(x,y)$. **c**, $B_z(x,y)$ measured by the SOT above the WTe$_2$ sample under current $I_0 = 50$ μA at 4.5 K. **d**, $J_x(x,y)$ reconstructed from (**c**). **e**, Simulated $J_x(x,y)$. **f**, Simulated current streamlines showing laminar flow. **g-l,** Measurements of WTe$_2$ sample with $\theta = 20°$ and corresponding simulations in the hydrodynamic regime. **g**, Current density $J_y(x,y)$ reconstructed from (**i**). See Extended Data Figs. 2k-o in which the color scale is expanded so that the counterflow vortex current $J_y(x,y)$ is resolved. **h**, Simulated $J_y(x,y)$. **i**, $B_z(x,y)$ measured in the WTe$_2$ sample carrying $I_0 = 50$ μA. **j**, $J_x(x,y)$ reconstructed from (**l**). **k**, Simulated $J_x(x,y)$. **l**, Simulated current flow showing laminar (open-loop, red) and vortex (closed-loop, blue) streamlines.

### Hydrodynamic and ballistic vortex stability phase diagram

Though the above hydrodynamic simulations show a good agreement with the experimental data, a question arises whether ballistic trajectories may create similar vortex patterns. Indeed, several studies have pointed out the difficulty in distinguishing the hydrodynamic and ballistic regimes using transport data [3,23,40,46,59]. Moreover, recent studies of current profiles in a WTe$_2$ whisker suggest the transport should be ballistic at low temperatures [10]. A key aspect that enables formation of vortices in both the hydrodynamic and ballistic regimes is system geometry [36]. Another key aspect is the boundary conditions, which has been an outstanding question in electron hydrodynamics from its early days [1,7–10,14,17,18,34,36,45,63,64,67]. The boundary conditions can be parametrized by the slip length, $\xi$, which can vary from $\xi = 0$ for no-slip ($\mathbf{J}|_{\text{boundary}} = 0$) to $\xi = \infty$ for no-stress boundaries (free-surface, $\hat{n} \cdot \nabla \mathbf{J}|_{\text{boundary}} = 0$). A full treatment of the electron transport requires the use of the Boltzmann kinetic equation, solution of which in an arbitrary 2D geometry is quite challenging [6,58]. We note, however, that for realistic parameters, Eq. 1 provides a good approximation in the regime of interest, $D/W \lesssim 1$, as well as in the quasi-ballistic regime, $D/W \gtrsim 1$, provided the no-stress boundary conditions are used [46] (see Methods).

To model vortex formation, we solve Eq. 1 in the two-chamber geometry and compute the total counterflow current $I_v$ carried by the vortex, $I_v = \int_{-w/2}^{w/2} \left( |J_y(x,0)| - J_y(x,0) \right) \mathrm{d}x/4$, where $w = W + 2R[1 + \cos(\theta/2)]$ is the width of the structure in its widest section. Figures 3a,b show the resulting vortex stability phase diagram as a function of the aperture angle $\theta$ and the ratio of the Gurzhi length to the strip width, $D/W$, for no-stress and no-slip boundary conditions (see Methods). The resulting phase diagrams are quite similar, predicting that in the quasi-ballistic regime $D/W \gtrsim 1$ the vortices feature large counterflow $I_v$, are stable up to large angles $\theta$, and show weak dependence on $D/W$. In the hydrodynamic regime, to the contrary, the counterflow $I_v$ is lower, the presence of vortices is limited to smaller $\theta$, and the vortex stability is strongly dependent on $D/W$. At low $D/W$, the vortex-to-laminar (no-vortex) phase transition line $\theta_t(D/W)$ is linear (dashed green line) and is essentially independent of the boundary conditions. Yet, the total circulating current $I_v$ carried by the vortex is strongly dependent on the boundary conditions, showing about four-fold suppression of the maximum $I_v$ for no-slip boundaries due to enhanced momentum relaxation at the edges.

These insights imply that by fabricating a series of samples with varying $\theta$, one can pinpoint the transition angle $\theta_t$. This would allow one to *i*) determine whether the flow is ballistic (large $\theta_t$) or hydrodynamic (small $\theta_t$), *ii*) evaluate the maximum $I_v$, and *iii*) extract the value of $D$ in the hydrodynamic case. Moreover, the derived $I_v$ can provide an estimate of the electron slip length $\xi$. Accordingly, we have fabricated six samples from a single WTe$_2$ flake with a sequence of aperture angles $\theta = 20°, 35°, 54°, 72°, 90°,$ and $120°$ (Extended Data Fig. 6a). The observed current flow patterns are presented in Figs. 2, 3, and 4 (see Methods and Extended Data Figs. 7, 8, and 9 for additional samples and geometries). The six data points are overlaid on the two phase diagrams in Figs. 3a,b. A single vortex in each chamber is observed in two samples with the smallest $\theta$ (marked by ⊙), a double-vortex is found in the $\theta = 54°$ sample (marked by ⊖), while laminar (no-vortex) flow is found for the three largest $\theta$ (marked by ×). As described below, the $\theta = 54°$ sample resides very close to the phase transition line, which allows us to identify the transition angle $\theta_t \cong 54°$. The small value of $\theta_t$ clearly establishes the hydrodynamic nature of the observed current vortices.

**Gurzhi length and boundary conditions**

The obtained $\theta_t$ value translates into $D/W \cong 0.28$, which for $W = 550$ nm in our devices results in the Gurzhi length $D \cong 155$ nm, a value nearly independent of the boundary conditions type in this range of parameters. The boundary conditions, however, strongly impact the vortex current $I_v$ as seen in Figs. 3a,b and demonstrated in Figs. 3c-j. The experimentally derived $J_x(x, y)$ in $\theta = 20°$ and $35°$ samples are shown in Figs. 3c,g alongside simulated $J_x(x, y)$ for three values of the slip length, $\xi = 0, 200$ nm, and $\infty$. Figures 3f,j demonstrate that for no-slip boundary conditions, the intensity of the circulating current in the vortex is much weaker than the one measured experimentally. Circulating currents comparable to the experimental values in Figs. 3c,g can be achieved only for large $\xi \gtrsim 200$ nm, as shown in Figs. 3e,i. In this limit, the resulting current distribution is nearly identical to the one found for the no-stress boundary conditions, $\xi = \infty$, in Figs. 3d,h. The relatively low values of the measured $J_x/(I_0/W) \cong 0.02$, as compared to the significantly higher maximum values calculated for the ballistic regime, provide additional evidence for the hydrodynamic nature of the observed vortices.

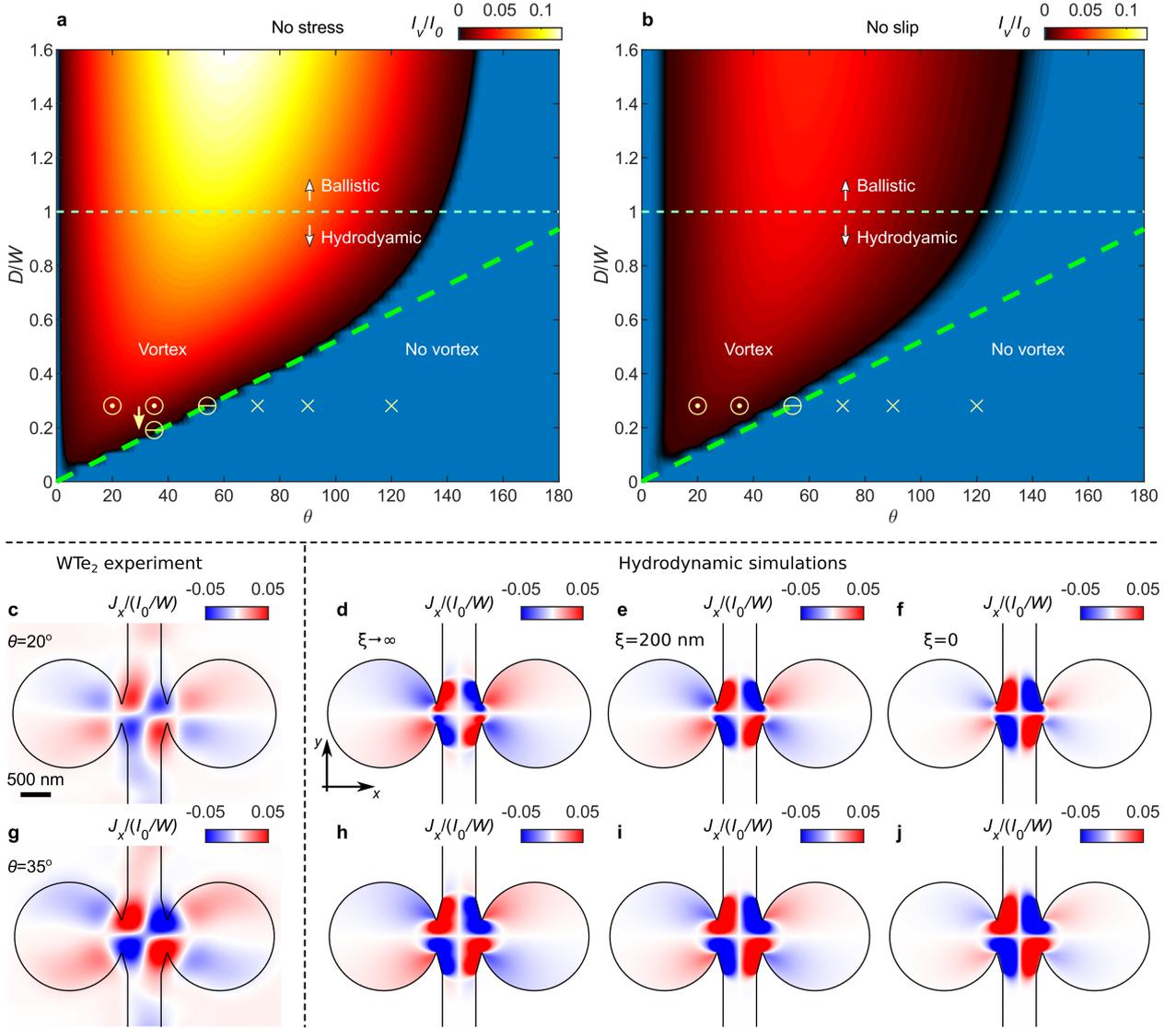

**Fig. 3**. **Vortex stability phase diagram. a-b**, Vortex stability phase diagram showing the magnitude of the circulating vortex current, $I_v$, in the chambers vs. the aperture angle $\theta$ and the Gurzhi length scaled by the strip width, $D/W$, for no-stress, $\xi = \infty$ (**a**), and no-slip, $\xi = 0$ (**b**), boundary conditions. The dashed green line indicates the vortical-to-laminar phase transition line $\theta_t(\frac{D}{W})$ in the $D/W \ll 1$ limit. The symbols $\odot$, $\ominus$, and × mark the parameters of the chambers that feature single vortex, double vortex, and no vortices, respectively. The double-vortex state at $\theta = 35°$ is described in Fig. 5. **c-f**, Measured $J_x(x,y)$ in WTe$_2$ sample with $\theta = 20°$ (**c**) and the corresponding simulated $J_y(x,y)$ for $D/W = 0.28$ and electron slip length at the edges of $\xi = \infty$ (**d**), 200 nm (**e**), and 0 (**f**). **g-j**, Measured $J_y(x,y)$ in WTe$_2$ sample with $\theta = 35°$ (**g**) and the corresponding simulated $J_x(x,y)$ for $\xi = \infty$ (**h**), 200 nm (**i**), and 0 (**j**).

Our finding of a large slip length $\xi \gtrsim 200$ nm in the hydrodynamic flow is consistent with several transport studies of graphene [1,34,36], but are in an apparent disagreement with recent spatially resolved studies of graphene [7–9] and WTe$_2$ [10], which have suggested diffuse or no-slip boundary conditions in the hydrodynamic regime. Note, however, that these studies are based on the analysis of current profiles in a strip geometry in which ballistic transport and hydrodynamic flow with large slip length result in essentially indistinguishable current profiles (see further discussion in Methods and Extended Data Fig. 3).

**Transition from laminar to vortex flow**

We now examine more closely the transition between laminar and vortex flows. Figure 4 shows the experimental current distributions in samples with apertures $\theta = 90°$, $72°$, $54°$, and $35°$. The first two geometries, $\theta = 90°$ and $72°$, show laminar (no vortex) flow, in good agreement with the numerical results (Figs. 4a-j). In the sample with a smaller aperture $\theta = 35°$, a well-resolved vortex is observed in each chamber—also in agreement with simulations (Figs. 4p-t). The intermediate $\theta = 54°$ aperture shows, however, a markedly different $J_x(x, y)$ flow pattern (Fig. 4l). The corresponding numerical simulations (Fig. 4m-o) reveal that at the vortex-to-no-vortex phase transition ($\theta = \theta_t$), rather than vanishing continuously, the vortex elongates into an arc (see Methods and Supplementary Video 1) and eventually splits into two sub-vortices in the top and bottom parts of the chamber as shown by the streamlines in Fig. 4o. As a result, in each chamber, $J_x(x, y)$ shows two pairs of red-blue streaks, one for each sub-vortex (Fig. 4l), instead of a single pair of streaks as in Fig. 4q. The numerical simulations show that the double-vortex flow occurs only in a narrow interval of parameters just below the $\theta_t(D/W)$ phase transition line (where $I_v = 0$ according to our definition, because all the streamlines at $y = 0$ become laminar), which allows us to determine $\theta_t \cong 54°$ and $D/W \cong 0.28$ in our devices. Importantly, the double-vortex state can occur only in the hydrodynamic regime and is precluded in the ballistic transport (see Methods and Supplementary Video 2), thus providing additional strong evidence for the hydrodynamic origin of the observed vortex flow.

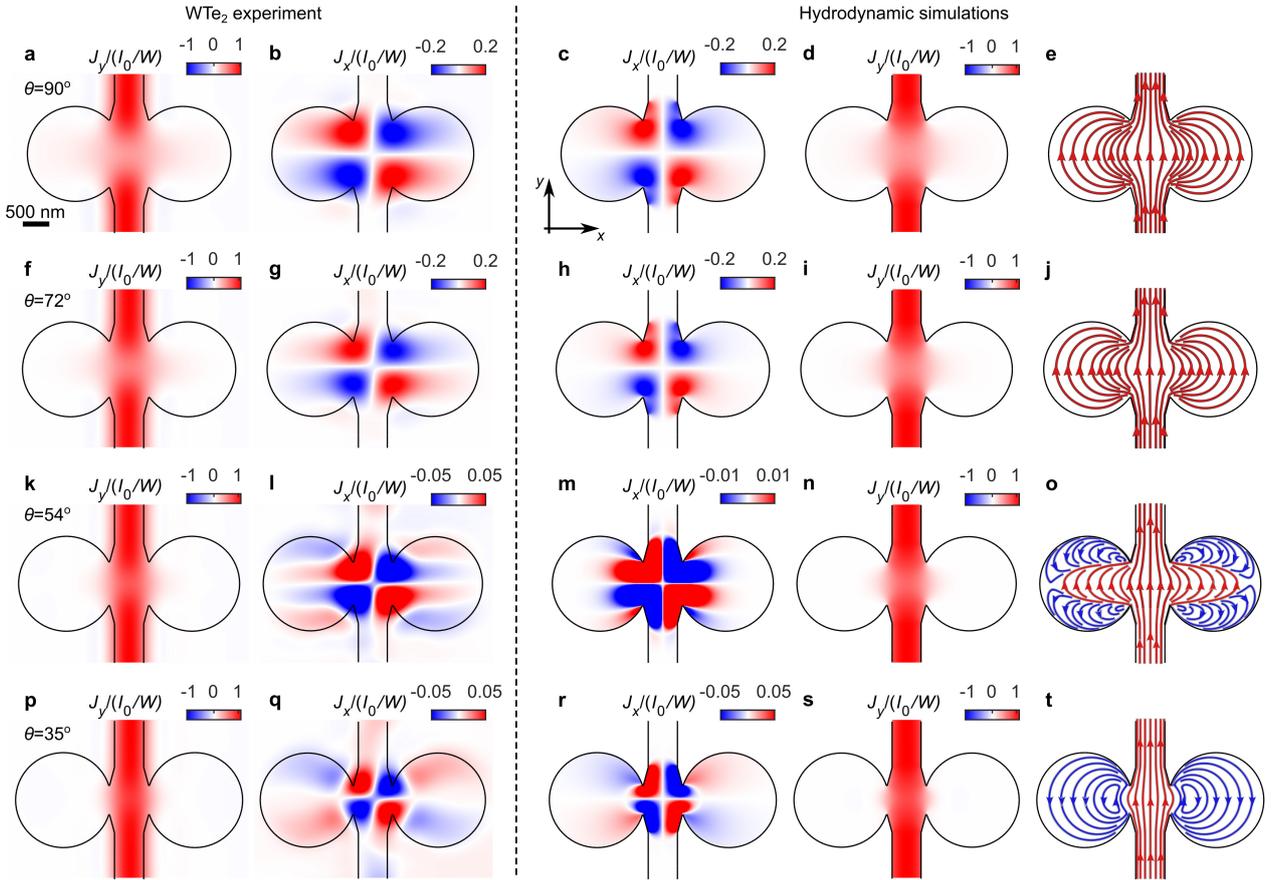

**Fig. 4**. **Laminar to double-vortex to single-vortex transition. a-e**, Measurements of WTe$_2$ sample with $\theta = 90°$ and corresponding simulations in the hydrodynamic regime with $D = 155$ nm and $\xi = 200$ nm. **a**, Measured current density $J_y(x, y)$ normalized by $I_0/W$ at $I_0 = 50$ μA. **b**, Measured current density $J_x(x, y)$. **c**, Simulated $J_x(x, y)$. **d**, Simulated $J_y(x, y)$. **e**, Simulated laminar (red) current streamlines. **f-j**, Same as (a-e) for $\theta = 72°$ sample. **k-o**, Same as (a-e) for $\theta = 54°$ sample showing double vortex with laminar (red) and vortex (blue) current streamlines in (o). **p-t**, Same as (a-e) for $\theta = 35°$ sample showing single vortex with laminar (red) and vortex (blue) current streamlines in (t).

## Current dependence

Our SOT microscope setup is limited to operation at temperatures of about 4 K, but the electron temperature can be raised substantially by increasing the applied current, as was originally employed in the hydrodynamic studies in GaAs 2DEG [17,18]. Since both $l_{ee}$ and $l_{mr}$ decrease with temperature, increasing the current $I_0$ is expected to reduce the Gurzhi length $D$. Figure 5 shows the evolution of $J_x(x, y)$ in the $\theta = 35°$ sample upon increasing $I_0$ up to 400 µA. The normalized $J_x/(I_0/W)$ circulating in the vortex gradually decreases as $I_0$ grows, until the double-vortex state is formed at our highest applied current. This behavior is qualitatively consistent with the expected decrease of $D$ as indicated by the arrow in Fig. 3a. The degree of the reduction in $D$ is, however, surprising. By measuring the sample resistance as a function of $I_0$ and comparing it to the temperature dependence of the resistance, we infer that the electron temperature reaches about 18 K at our highest applied current. Based on the theoretical estimate of the temperature dependence of $l_{ee}$ and $l_{mr}$ [10], such temperature increase should have reduced $D$ by more than an order of magnitude relative to the value at 4.5 K. In contrast, Fig. 3a indicates that $D$ has decreased by less than a factor of two. Note that in Ref. [10], the measured temperature evolution was also found to be weaker than predicted. This finding of weak temperature dependence of the Gurzhi length provides an important insight into the mechanism underlying hydrodynamics in our system.

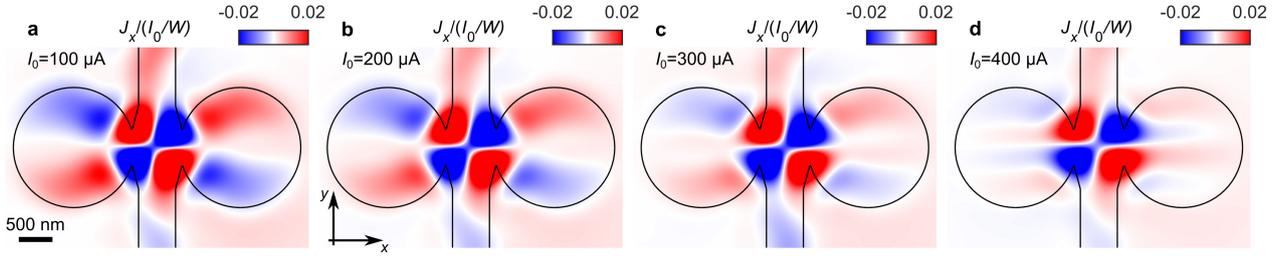

**Fig. 5**. **Current dependence of the vortex state. a-d**, $J_x(x, y)$ measured in WTe₂ sample with $\theta = 35°$ normalized by $I_0/W$ for applied currents of $I_0 = 100$ µA (**a**), 200 µA (**b**), 300 µA (**c**), and 400 µA (**d**), showing a transition from single-vortex to double-vortex state in the chambers.

## Discussion

Perhaps our most unexpected finding is that we observe fluid-like spatial momentum transport characterized by $D \cong 155$ nm, a value that is much smaller than $D = \sqrt{l_{mr}l_{ee}}/2$ estimated based on the bulk microscopic parameters. Indeed, transport measurements indicate that $l_{mr}$ values in our bulk samples are in excess of 10 µm (Methods and Extended Data Fig. 1). Recent band structure calculations [10] suggest that bare $l_{ee}$ in WTe₂ is of the order of a few mm at 4 K. Accounting for phonon-mediated electron-electron interactions can reduce $l_{ee}$ to about 100 µm [10]. By taking into account the compensated semimetal band structure of WTe₂ and the proximity to band edges, we show that the bare $l_{ee}$ can be of the order of 10 µm (Methods and Extended Data Fig. 5). Yet, even with this lower bound, $l_{ee}$ remains much larger than $W$, which should have resulted in transport deep in the ballistic regime, with $D = \sqrt{l_{mr}l_{ee}}/2 \cong 5$ µm.

To gain insight into the origin of this surprising behavior, it is instructive to recall the general derivation of the Ohm-Stokes law, Eq. 1. Kinetic theory links momentum relaxation to the decay rate $\gamma_1$ of the first angular harmonic of the nonequilibrium electron momentum distribution, giving $l_{mr} = v_F/\gamma_1$, whereas the kinematic viscosity $\eta = \frac{1}{4}\frac{v_F^2}{\gamma_2}$ is expressed through the decay rate $\gamma_2$ of the second harmonic of the momentum distribution, where $v_F$ is the Fermi velocity (Methods). This results in $D = \sqrt{\eta/\gamma_1} = v_F/\sqrt{4\gamma_1\gamma_2}$ (see also [35]). Importantly, this expression is completely general and is valid for any microscopic momentum-scattering mechanism. Usually $\gamma_1$ and $\gamma_2$ originate, respectively, from impurity or phonon scattering and electron-electron collisions, giving $\gamma_1 = v_F/l_{mr}$, $\gamma_2 = v_F/l_{ee}$, and resulting in $D =$ $\sqrt{l_{mr}l_{ee}}/2$. As discussed above, because of the large values of bulk $l_{mr}$ and $l_{ee}$, these relations are inconsistent with $D$ inferred from the observed behavior.

However, momentum-conserving electron-electron scattering is not required in order to generate spatial diffusion of the momentum. There is another scattering mechanism that could result in both the diffusion and relaxation of electron momenta, related to the finite thickness $d$ of the sample. The resistivity of WTe$_2$ flakes has been found to be strongly thickness dependent [68,69], a behavior attributed to surface oxidation. Our transport measurements show that the conductivity of the flakes is one to two orders of magnitude lower than that of the bulk crystals (see Methods), indicating a large reduction in $l_{mr}$ and an enhancement of $\gamma_1$ induced by the momentum relaxing scattering off the surfaces. However, an enhancement of $\gamma_1$ alone, such that $\gamma_1 > \gamma_2$, would of course lead to an ohmic transport with no vorticity, in contrast to the observed hydrodynamic flow, which requires $\gamma_2 > \gamma_1$. Yet, it has recently been pointed out [35] that enhancement in $\gamma_1$ implies an enhancement in $\gamma_2$. Indeed, using the method described in [59], we show in Methods that small-angle scattering results in $\gamma_2 \cong 4\gamma_1$, giving rise to hydrodynamic-like transport with $D = v_F/\sqrt{4\gamma_1\gamma_2} = v_F/4\gamma_1 = l_{mr}/4$. Based on $D \cong 155$ nm derived from the vortex stability diagram, we arrive at the effective surface-induced $l_{mr} = 4D \cong 620$ nm in our samples. This value compares well with the effective $l_{mr} \cong 530$ nm derived independently from transport measurements of conductivity in thin flakes (Methods).

The emerging picture is therefore as follows. For a fully specular surface scattering, the transport is ballistic. Small-angle scattering at the surfaces results in two effects: enhancement in momentum relaxation and a concurrent enhancement of the lateral momentum diffusion. In this para-hydrodynamic mechanism, momentum diffusion does not occur through the usual momentum-conserving electron-electron scattering, but rather through multiple, close-to-specular scattering events between the top and bottom surfaces without inter-electron momentum transfer. The resulting surface-induced para-hydrodynamics opens a unique possibility of observing and utilizing hydrodynamic phenomena in a wide range of high mobility materials without the necessity of the hard-to-achieve strong bulk electron-electron interactions.

**References**

1. D. A. Bandurin, I. Torre, R. K. Kumar, M. Ben Shalom, A. Tomadin, A. Principi, G. H. Auton, E. Khestanova, K. S. Novoselov, I. V Grigorieva, L. A. Ponomarenko, A. K. Geim, and M. Polini, ''Negative local resistance caused by viscous electron backflow in graphene'', *Science* **351**, 1055–1058 (2016).

2. A. D. Levin, G. M. Gusev, E. V. Levinson, Z. D. Kvon, and A. K. Bakarov, ''Vorticity-induced negative nonlocal resistance in a viscous two-dimensional electron system'', *Phys. Rev. B* **97**, 245308 (2018).

3. D. A. Bandurin, A. V. Shytov, L. S. Levitov, R. K. Kumar, A. I. Berdyugin, M. Ben Shalom, I. V. Grigorieva, A. K. Geim, and G. Falkovich, ''Fluidity onset in graphene'', *Nat. Commun.* **9**, 4533 (2018).

4. A. I. Berdyugin, S. G. Xu, F. M. D. Pellegrino, R. Krishna Kumar, A. Principi, I. Torre, M. Ben Shalom, T. Taniguchi, K. Watanabe, I. V. Grigorieva, M. Polini, A. K. Geim, and D. A. Bandurin, ''Measuring Hall viscosity of graphene's electron fluid'', *Science* **364**, 162–165 (2019).

5. M. Kim, S. G. Xu, A. I. Berdyugin, A. Principi, S. Slizovskiy, N. Xin, P. Kumaravadivel, W. Kuang, M. Hamer, R. Krishna Kumar, R. V. Gorbachev, K. Watanabe, T. Taniguchi, I. V. Grigorieva, V. I. Fal'ko, M. Polini, and A. K. Geim, ''Control of electron-electron interaction in graphene by proximity screening'', *Nat. Commun.* **11**, 2339 (2020).

6. A. Gupta, J. J. Heremans, G. Kataria, M. Chandra, S. Fallahi, G. C. Gardner, and M. J. Manfra, ''Hydrodynamic and Ballistic Transport over Large Length Scales in GaAs/AlGaAs'', *Phys. Rev. Lett.* **126**, 076803 (2021).

7. J. A. Sulpizio, L. Ella, A. Rozen, J. Birkbeck, D. J. Perello, D. Dutta, M. Ben-Shalom, T. Taniguchi, K. Watanabe, T. Holder, R. Queiroz, A. Principi, A. Stern, T. Scaffidi, A. K. Geim, and S. Ilani, ''Visualizing Poiseuille flow of hydrodynamic electrons'', *Nature* **576**, 75–79 (2019).

8. M. J. H. Ku, T. X. Zhou, Q. Li, Y. J. Shin, J. K. Shi, C. Burch, L. E. Anderson, A. T. Pierce, Y. Xie, A. Hamo, U. Vool, H. Zhang, F. Casola, T. Taniguchi, K. Watanabe, M. M. Fogler, P. Kim, A. Yacoby, and R. L. Walsworth, ''Imaging viscous flow of the Dirac fluid in graphene'', *Nature* **583**, 537–541 (2020).

9. A. Jenkins, S. Baumann, H. Zhou, S. A. Meynell, D. Yang, K. Watanabe, T. Taniguchi, A. Lucas, A. F. Young, and A. C. B. Jayich, ''Imaging the breakdown of ohmic transport in graphene'', *arXiv:2002.05065* (2020).

10. U. Vool, A. Hamo, G. Varnavides, Y. Wang, T. X. Zhou, N. Kumar, Y. Dovzhenko, Z. Qiu, C. A. C. Garcia, A. T. Pierce, J. Gooth, P. Anikeeva, C. Felser, P. Narang, and A. Yacoby, ''Imaging phonon-mediated hydrodynamic flow in WTe2'', *Nat. Phys.* **17**, 1216–1220 (2021).

11. J. Crossno, J. K. Shi, K. Wang, X. Liu, A. Harzheim, A. Lucas, S. Sachdev, P. Kim, T. Taniguchi, K. Watanabe, T. A. Ohki, and K. C. Fong, ''Observation of the Dirac fluid and the breakdown of the Wiedemann-Franz law in graphene'', *Science* **351**, 1058–1061 (2016).

12. J. Gooth, F. Menges, N. Kumar, V. Süβ, C. Shekhar, Y. Sun, U. Drechsler, R. Zierold, C. Felser, and B. Gotsmann, ''Thermal and electrical signatures of a hydrodynamic electron fluid in tungsten diphosphide'', *Nat. Commun.* **9**, 4093 (2018).

13. D. Vasyukov, Y. Anahory, L. Embon, D. Halbertal, J. Cuppens, L. Neeman, A. Finkler, Y. Segev, Y. Myasoedov, M. L. Rappaport, M. E. Huber, and E. Zeldov, ''A scanning superconducting quantum interference device with single electron spin sensitivity'', *Nat. Nanotechnol.* **8**, 639–644 (2013).

14. R. N. Gurzhi, ''Hydrodynamic effects in solids at low temperature'', *Sov. Phys. Uspekhi* **11**, 255–270 (1968).

15. L. D. Landau and E. M. Lifshitz, *Fluid Mechanics*. Elsevier, 1987.

16. J. Mayzel, V. Steinberg, and A. Varshney, ''Stokes flow analogous to viscous electron current in graphene'', *Nat. Commun.* **10**, 937 (2019).

17. L. W. Molenkamp and M. J. M. de Jong, ''Observation of Knudsen and Gurzhi transport regimes in a two-dimensional wire'', *Solid. State. Electron.* **37**, 551–553 (1994).

18. M. J. M. de Jong and L. W. Molenkamp, ''Hydrodynamic electron flow in high-mobility wires'', *Phys. Rev. B* **51**, 13389–13402 (1995).

19. O. E. Raichev, G. M. Gusev, A. D. Levin, and A. K. Bakarov, ''Manifestations of classical size effect and electronic viscosity in the magnetoresistance of narrow two-dimensional conductors: Theory and experiment'', *Phys. Rev. B* **101**, 235314 (2020).

20. D. Taubert, G. J. Schinner, C. Tomaras, H. P. Tranitz, W. Wegscheider, and S. Ludwig, ''An electron jet pump: The venturi effect of a fermi liquid'', *J. Appl. Phys.* **109**, (2011).

21. P. J. W. Moll, P. Kushwaha, N. Nandi, B. Schmidt, and A. P. Mackenzie, ''Evidence for hydrodynamic electron flow in PdCoO2'', *Science* **351**, 1061–1064 (2016).

22. R. Krishna Kumar, D. A. Bandurin, F. M. D. Pellegrino, Y. Cao, A. Principi, H. Guo, G. H. Auton, M. Ben Shalom, L. A. Ponomarenko, G. Falkovich, K. Watanabe, T. Taniguchi, I. V. Grigorieva, L. S. Levitov, M. Polini, and A. K. Geim, ''Superballistic flow of viscous electron fluid through graphene constrictions'', *Nat. Phys.* **13**, 1182–1185 (2017).

23. B. A. Braem, F. M. D. Pellegrino, A. Principi, M. Röösli, C. Gold, S. Hennel, J. V. Koski, M. Berl, W. Dietsche, W. Wegscheider, M. Polini, T. Ihn, and K. Ensslin, ''Scanning gate microscopy in a viscous electron fluid'', *Phys. Rev. B* **98**, 241304 (2018).

24. G. M. Gusev, A. S. Jaroshevich, A. D. Levin, Z. D. Kvon, and A. K. Bakarov, "Stokes flow around an obstacle in viscous two-dimensional electron liquid", *Sci. Rep.* **10**, 7860 (2020).

25. G. M. Gusev, A. S. Jaroshevich, A. D. Levin, Z. D. Kvon, and A. K. Bakarov, "Viscous magnetotransport and Gurzhi effect in bilayer electron system", *Phys. Rev. B* **103**, 075303 (2021).

26. Z. J. Krebs, W. A. Behn, S. Li, K. J. Smith, K. Watanabe, T. Taniguchi, A. Levchenko, and V. W. Brar, "Imaging the breaking of electrostatic dams in graphene for ballistic and viscous fluids", *arXiv:2106.07212* (2021).

27. J. Geurs, Y. Kim, K. Watanabe, T. Taniguchi, P. Moon, and J. H. Smet, "Rectification by hydrodynamic flow in an encapsulated graphene Tesla valve", *arXiv:2008.04862* (2020).

28. S. Samaddar, J. Strasdas, K. Janßen, S. Just, T. Johnsen, Z. Wang, B. Uzlu, S. Li, D. Neumaier, M. Liebmann, and M. Morgenstern, "Evidence for Local Spots of Viscous Electron Flow in Graphene at Moderate Mobility", *Nano Lett.* **21**, 9365–9373 (2021).

29. C. Kumar, J. Birkbeck, J. A. Sulpizio, D. J. Perello, T. Taniguchi, K. Watanabe, O. Reuven, T. Scaffidi, A. Stern, A. K. Geim, and S. Ilani, "Imaging Hydrodynamic Electrons Flowing Without Landauer-Sharvin Resistance", *arXiv:2111.06412* (2021).

30. A. O. Govorov and J. J. Heremans, "Hydrodynamic Effects in Interacting Fermi Electron Jets", *Phys. Rev. Lett.* **92**, 026803 (2004).

31. M. Müller, J. Schmalian, and L. Fritz, "Graphene: A Nearly Perfect Fluid", *Phys. Rev. Lett.* **103**, 2–5 (2009).

32. A. V. Andreev, S. A. Kivelson, and B. Spivak, "Hydrodynamic Description of Transport in Strongly Correlated Electron Systems", *Phys. Rev. Lett.* **106**, 256804 (2011).

33. M. Mendoza, H. J. Herrmann, and S. Succi, "Preturbulent Regimes in Graphene Flow", *Phys. Rev. Lett.* **106**, 156601 (2011).

34. I. Torre, A. Tomadin, A. K. Geim, and M. Polini, "Nonlocal transport and the hydrodynamic shear viscosity in graphene", *Phys. Rev. B* **92**, 165433 (2015).

35. P. S. Alekseev, "Negative Magnetoresistance in Viscous Flow of Two-Dimensional Electrons", *Phys. Rev. Lett.* **117**, 166601 (2016).

36. F. M. D. Pellegrino, I. Torre, A. K. Geim, and M. Polini, "Electron hydrodynamics dilemma: Whirlpools or no whirlpools", *Phys. Rev. B* **94**, 155414 (2016).

37. H. Guo, E. Ilseven, G. Falkovich, and L. S. Levitov, "Higher-than-ballistic conduction of viscous electron flows", *PNAS* **114**, 3068–3073 (2017).

38. T. Scaffidi, N. Nandi, B. Schmidt, A. P. Mackenzie, and J. E. Moore, "Hydrodynamic Electron Flow and Hall Viscosity", *Phys. Rev. Lett.* **118**, 226601 (2017).

39. V. Galitski, M. Kargarian, and S. Syzranov, "Dynamo Effect and Turbulence in Hydrodynamic Weyl Metals", *Phys. Rev. Lett.* **121**, 176603 (2018).

40. A. Shytov, J. F. Kong, G. Falkovich, and L. Levitov, "Particle Collisions and Negative Nonlocal Response of Ballistic Electrons", *Phys. Rev. Lett.* **121**, 176805 (2018).

41. D. Svintsov, "Hydrodynamic-to-ballistic crossover in Dirac materials", *Phys. Rev. B* **97**, 121405 (2018).

42. P. S. Alekseev, A. P. Dmitriev, I. V. Gornyi, V. Y. Kachorovskii, B. N. Narozhny, and M. Titov, "Nonmonotonic magnetoresistance of a two-dimensional viscous electron-hole fluid in a confined geometry", *Phys. Rev. B* **97**, 085109 (2018).

43. P. S. Alekseev, A. P. Dmitriev, I. V. Gornyi, V. Y. Kachorovskii, B. N. Narozhny, and M. Titov, "Counterflows in viscous electron-hole fluid", *Phys. Rev. B* **98**, 125111 (2018).

44. I. S. Burmistrov, M. Goldstein, M. Kot, V. D. Kurilovich, and P. D. Kurilovich, "Dissipative and Hall Viscosity of a Disordered 2D Electron Gas", *Phys. Rev. Lett.* **123**, 26804 (2019).

45. K. A. Guerrero-Becerra, F. M. D. Pellegrino, and M. Polini, "Magnetic hallmarks of viscous electron

flow in graphene'', *Phys. Rev. B* **99**, 041407 (2019).

46. T. Holder, R. Queiroz, T. Scaffidi, N. Silberstein, A. Rozen, J. A. Sulpizio, L. Ella, S. Ilani, and A. Stern, ''Ballistic and hydrodynamic magnetotransport in narrow channels'', *Phys. Rev. B* **100**, 245305 (2019).

47. P. Ledwith, H. Guo, A. Shytov, and L. Levitov, ''Tomographic Dynamics and Scale-Dependent Viscosity in 2D Electron Systems'', *Phys. Rev. Lett.* **123**, 116601 (2019).

48. B. N. Narozhny and M. Schütt, ''Magnetohydrodynamics in graphene: Shear and Hall viscosities'', *Phys. Rev. B* **100**, 035125 (2019).

49. D. Di Sante, J. Erdmenger, M. Greiter, I. Matthaiakakis, R. Meyer, D. R. Fernández, R. Thomale, E. van Loon, and T. Wehling, ''Turbulent hydrodynamics in strongly correlated Kagome metals'', *Nat. Commun.* **11**, 3997 (2020).

50. R. Toshio, K. Takasan, and N. Kawakami, ''Anomalous hydrodynamic transport in interacting noncentrosymmetric metals'', *Phys. Rev. Res.* **2**, 032021 (2020).

51. E. H. Hasdeo, J. Ekström, E. G. Idrisov, and T. L. Schmidt, ''Electron hydrodynamics of two-dimensional anomalous Hall materials'', *Phys. Rev. B* **103**, 125106 (2021).

52. Y. Huang and M. Wang, ''Nonnegative magnetoresistance in hydrodynamic regime of electron fluid transport in two-dimensional materials'', *Phys. Rev. B* **104**, 155408 (2021).

53. A. Hui, V. Oganesyan, and E. Kim, ''Beyond Ohm's law: Bernoulli effect and streaming in electron hydrodynamics'', *Phys. Rev. B* **103**, 235152 (2021).

54. B. N. Narozhny, I. V. Gornyi, and M. Titov, ''Anti-Poiseuille flow in neutral graphene'', *Phys. Rev. B* **104**, 075443 (2021).

55. B. N. Narozhny, I. V. Gornyi, and M. Titov, ''Hydrodynamic collective modes in graphene'', *Phys. Rev. B* **103**, 115402 (2021).

56. O. Tavakol and Y. B. Kim, ''Artificial electric field and electron hydrodynamics'', *Phys. Rev. Res.* **3**, 013290 (2021).

57. G. Zhang, V. Kachorovskii, K. Tikhonov, and I. Gornyi, ''Heating of inhomogeneous electron flow in the hydrodynamic regime'', *Phys. Rev. B* **104**, 075417 (2021).

58. S. Li, M. Khodas, and A. Levchenko, ''Conformal maps of viscous electron flow in the Gurzhi crossover'', *Phys. Rev. B* **104**, 155305 (2021).

59. K. G. Nazaryan and L. Levitov, ''Robustness of vorticity in electron fluids'', *arXiv:2111.09878* (2021).

60. A. Principi, G. Vignale, M. Carrega, and M. Polini, ''Bulk and shear viscosities of the two-dimensional electron liquid in a doped graphene sheet'', *Phys. Rev. B* **93**, 125410 (2016).

61. A. Stern, T. Scaffidi, O. Reuven, C. Kumar, J. Birkbeck, and S. Ilani, ''Spread and erase -- How electron hydrodynamics can eliminate the Landauer-Sharvin resistance'', *arXiv:2110.15369* (2021).

62. L. Levitov and G. Falkovich, ''Electron viscosity, current vortices and negative nonlocal resistance in graphene'', *Nat. Phys.* **12**, 672–676 (2016).

63. G. Falkovich and L. Levitov, ''Linking Spatial Distributions of Potential and Current in Viscous Electronics'', *Phys. Rev. Lett.* **119**, 066601 (2017).

64. S. Danz and B. N. Narozhny, ''Vorticity of viscous electronic flow in graphene'', *2D Mater.* **7**, 035001 (2020).

65. A. Gabbana, M. Polini, S. Succi, R. Tripiccione, and F. M. D. Pellegrino, ''Prospects for the Detection of Electronic Preturbulence in Graphene'', *Phys. Rev. Lett.* **121**, 236602 (2018).

66. A. Y. Meltzer, E. Levin, and E. Zeldov, ''Direct Reconstruction of Two-Dimensional Currents in Thin Films from Magnetic-Field Measurements'', *Phys. Rev. Appl.* **8**, 064030 (2017).

67. E. I. Kiselev and J. Schmalian, ''Boundary conditions of viscous electron flow'', *Phys. Rev. B* **99**,

035430 (2019).

68. L. Wang, I. Gutiérrez-Lezama, C. Barreteau, N. Ubrig, E. Giannini, and A. F. Morpurgo, ''Tuning magnetotransport in a compensated semimetal at the atomic scale'', *Nat. Commun.* **6**, 8892 (2015).

69. J. M. Woods, J. Shen, P. Kumaravadivel, Y. Pang, Y. Xie, G. A. Pan, M. Li, E. I. Altman, L. Lu, and J. J. Cha, ''Suppression of Magnetoresistance in Thin WTe2 Flakes by Surface Oxidation'', *ACS Appl. Mater. Interfaces* **9**, 23175–23180 (2017).

70. M. N. Ali, J. Xiong, S. Flynn, J. Tao, Q. D. Gibson, L. M. Schoop, T. Liang, N. Haldolaarachchige, M. Hirschberger, N. P. Ong, and R. J. Cava, ''Large, non-saturating magnetoresistance in WTe2'', *Nature* **514**, 205–208 (2014).

71. Y.-Y. Lv, X. Li, B.-B. Zhang, W. Y. Deng, S.-H. Yao, Y. B. Chen, J. Zhou, S.-T. Zhang, M.-H. Lu, L. Zhang, M. Tian, L. Sheng, and Y.-F. Chen, ''Experimental Observation of Anisotropic Adler-Bell-Jackiw Anomaly in Type-II Weyl Semimetal WTe1.98 Crystals at the Quasiclassical Regime'', *Phys. Rev. Lett.* **118**, 096603 (2017).

72. Y. Wu, N. H. Jo, M. Ochi, L. Huang, D. Mou, S. L. Bud'ko, P. C. Canfield, N. Trivedi, R. Arita, and A. Kaminski, ''Temperature-Induced Lifshitz Transition in WTe2'', *Phys. Rev. Lett.* **115**, 166602 (2015).

73. M. N. Ali, L. Schoop, J. Xiong, S. Flynn, Q. Gibson, M. Hirschberger, N. P. Ong, and R. J. Cava, ''Correlation of crystal quality and extreme magnetoresistance of WTe2'', *EPL Europhys. Lett.* **110**, 67002 (2015).

74. Z. Zhu, X. Lin, J. Liu, B. Fauqué, Q. Tao, C. Yang, Y. Shi, and K. Behnia, ''Quantum Oscillations, Thermoelectric Coefficients, and the Fermi Surface of Semimetallic WTe2'', *Phys. Rev. Lett.* **114**, 176601 (2015).

75. F.-X. Xiang, M. Veldhorst, S.-X. Dou, and X.-L. Wang, ''Multiple Fermi pockets revealed by Shubnikov-de Haas oscillations in WTe2'', *EPL* **112**, 37009 (2015).

76. Q. Zhang, Z. Liu, Y. Sun, H. Yang, J. Jiang, S. Mo, Z. Hussain, X. Qian, L. Fu, S. Yao, M. Lu, C. Felser, B. Yan, Y. Chen, and L. Yang, ''Lifshitz Transitions Induced by Temperature and Surface Doping in Type-II Weyl Semimetal Candidate Td -WTe2'', *Phys. Status Solidi RRL* **11**, 1700209 (2017).

77. Y. Luo, H. Li, Y. M. Dai, H. Miao, Y. G. Shi, H. Ding, A. J. Taylor, D. A. Yarotski, R. P. Prasankumar, and J. D. Thompson, ''Hall effect in the extremely large magnetoresistance semimetal WTe2'', *Appl. Phys. Lett.* **107**, 182411 (2015).

78. P. Wang, G. Yu, Y. Jia, M. Onyszczak, F. A. Cevallos, S. Lei, S. Klemenz, K. Watanabe, T. Taniguchi, R. J. Cava, L. M. Schoop, and S. Wu, ''Landau quantization and highly mobile fermions in an insulator'', *Nature* **589**, 225–229 (2021).

79. N. Kumar, Y. Sun, N. Xu, K. Manna, M. Yao, V. Süss, I. Leermakers, O. Young, T. Förster, M. Schmidt, H. Borrmann, B. Yan, U. Zeitler, M. Shi, C. Felser, and C. Shekhar, ''Extremely high magnetoresistance and conductivity in the type-II Weyl semimetals WP2 and MoP2'', *Nat. Commun.* **8**, 1642 (2017).

80. Y. Anahory, H. R. Naren, E. O. Lachman, S. Buhbut Sinai, A. Uri, L. Embon, E. Yaakobi, Y. Myasoedov, M. E. Huber, R. Klajn, and E. Zeldov, ''SQUID-on-tip with single-electron spin sensitivity for high-field and ultra-low temperature nanomagnetic imaging'', *Nanoscale* **12**, 3174–3182 (2020).

81. M. E. Huber, P. A. Neil, R. G. Benson, D. A. Burns, A. M. Corey, C. S. Flynn, Y. Kitaygorodskaya, O. Massihzadeh, J. M. Martinis, and G. C. Hilton, ''DC SQUID series array amplifiers with 120 MHz bandwidth'', *IEEE Trans. Appiled Supercond.* **11**, 1251–1256 (2001).

82. A. Finkler, Y. Segev, Y. Myasoedov, M. L. Rappaport, L. Ne'eman, D. Vasyukov, E. Zeldov, M. E. Huber, J. Martin, and A. Yacoby, ''Self-Aligned Nanoscale SQUID on a Tip'', *Nano Lett.* **10**, 1046–1049 (2010).

83. A. Finkler, D. Vasyukov, Y. Segev, L. Ne'eman, E. O. Lachman, M. L. Rappaport, Y. Myasoedov, E. Zeldov, and M. E. Huber, ''Scanning superconducting quantum interference device on a tip for magnetic imaging of nanoscale phenomena", *Rev. Sci. Instrum.* **83**, 073702 (2012).

84. D. Halbertal, J. Cuppens, M. Ben Shalom, L. Embon, N. Shadmi, Y. Anahory, H. R. Naren, J. Sarkar, A. Uri, Y. Ronen, Y. Myasoedov, L. S. Levitov, E. Joselevich, A. K. Geim, and E. Zeldov, "Nanoscale thermal imaging of dissipation in quantum systems", *Nature* **539**, 407–410 (2016).

85. D. A. Broadway, S. E. Lillie, S. C. Scholten, D. Rohner, N. Dontschuk, P. Maletinsky, J.-P. Tetienne, and L. C. L. Hollenberg, "Improved Current Density and Magnetization Reconstruction Through Vector Magnetic Field Measurements", *Phys. Rev. Appl.* **14**, 024076 (2020).

86. L. Dell'Anna and W. Metzner, "Fermi surface fluctuations and single electron excitations near Pomeranchuk instability in two dimensions", *Phys. Rev. B* **73**, 045127 (2006).

**Acknowledgments** The authors thank Michal Shavit and Victor Steinberg for useful discussions. This work was supported by the European Research Council (ERC) under the European Union's Horizon 2020 research and innovation program (grant No. 785971), by the German-Israeli Foundation for Scientific Research and Development (GIF) grant No. I-1505-303.10/2019, and by the Israel Science Foundation (ISF) grant No. 994/19. G.F. was supported by the Scientific Excellence Center at WIS, the Simons Foundation grant 662962, the EU Horizon 2020 programme grant 873028, the U.S.-Israel Binational Science Foundation (BSF) grant 2018033 and NSF-BSF grant 2020765. B.Y. acknowledges the financial support by the European Research Council (ERC Consolidator Grant "NonlinearTopo", No. 815869) and the ISF grant No. 2932/21. E.Z. acknowledges the support of the Andre Deloro Prize for Scientific Research. L.S.V. and E.Z. acknowledge the support of the Sagol Weizmann-MIT Bridge Program. M.H. and E.Z. acknowledge the support of the Leona M. and Harry B. Helmsley Charitable Trust grants No. 2018PG-ISL006 and 2112-04911. A.K.P. acknowledges the postdoctoral fellowship support from the Council for Higher Education, Israel through 'Study in Israel' program.

**Author contributions** A.A., T.V., A.K. and E.Z. conceived the experiments. A.K.P. and M.H. grew and characterized the bulk $WTe_2$ crystals. A.K. fabricated and characterized the devices. T.V. and A.A. carried out the SOT magnetic imaging measurements and data analysis. I.R. and Y.M. fabricated the SOTs and the tuning fork feedback. A.Y.M. developed the current density reconstruction method. M.E.H. designed and built the SOT readout system. Y.W., T.H. and B.Y. carried out the band structure and the electron-electron scattering calculations. T.H. developed the vortex stability model. A.A. performed the numerical simulations. G.F., L.S.L. and E.Z. developed the para-hydrodynamic model. A.A., T.V., T.H., A.K.P, E.Z., G.F. and L.S.L. wrote the manuscript with contributions from the rest of the authors.

**Competing interests** The authors declare no competing interests.

**Data availability** The data that support the findings of this study are available from the corresponding authors on reasonable request.

**Code availability** The current reconstruction codes used in this study are available from the corresponding authors on reasonable request.

# Methods

## Synthesis of WTe$_2$ crystals

To obtain high quality WTe$_2$ single crystals we conducted a series of synthesis experiments using both chemical vapor transport (CVT) and the flux growth technique, as well as starting materials with different purity. These experiments led to progressively better crystals with increased $RRR = \frac{\rho(300\text{ K})}{\rho(2\text{ K})}$ and magneto resistance ratio $MR = \frac{\rho(9\text{ T}) - \rho(0\text{ T})}{\rho(0\text{ T})}$, as summarized in Extended Data Fig. 1c, and are described here in chronological order. Eventually all devices for the hydrodynamic flow experiments were fabricated from our highest quality single crystals described last. The first crystals were grown by CVT [70,71] using elemental W (99.95%) and Te (99.99%) from Stanford Advanced Materials. Initially, polycrystalline WTe$_2$ was prepared by solid-state reaction in a vacuum-sealed quartz ampule at 750 °C. The obtained precursor material was then vacuum-sealed (1.33 × 10$^{-5}$ mbar) in a 16 cm long quartz tube with a minute amount of TeBr$_4$ transport agent, and placed in a temperature gradient of 850 °C – 750 °C for several days. Few millimeters wide sheet-shaped single crystals were collected from the cold end of the ampule after cool down. However, we found that both $RRR$ and $MR$ of the CVT grown crystals were extremely low, despite testing a range of different growth conditions.

Much higher quality crystals were obtained using tellurium self-flux growth [72]. Elemental W and Te were mixed at a molar ratio of 1:30, loaded into frit-disc alumina crucibles, and sealed in a quartz ampule under vacuum. All steps of materials handling were performed in an Ar glove box with O$_2$ and H$_2$O concentration < 0.1 ppm. Quartz ampule, alumina crucibles, and quartz wool for cushioning were heat treated at 800 °C prior to the growth experiment. The tungsten tellurium mixture was heated in a box furnace to 1100 °C at a rate of 30 °C/h, followed by soaking at 1100 °C for 10 h. The metal solution was then slowly cooled down to 650 °C at a rate of 2 °C/h, followed by centrifuging to separate the Te flux from the crystals. To remove any trace of Te flux from the crystal surfaces, they were again vacuum sealed in a quartz ampule and placed on the hot side of a temperature gradient of 400 °C – 190 °C. Although these needle shaped crystals were of much higher quality than those grown by CVT, they were still inferior to the best crystals reported in literature [72]. Subsequent optimization of the growth parameters, such as a higher sample-to-flux ratio of W:Te = 1:50 and W:Te = 1:120, varying cooling rates, changing crucible arrangements, using pre-reacted WTe$_2$ and excess tellurium as starting materials, led to further improvements but the $RRR$ and $MR$ values were still unsatisfactory; see two points with lowest $RRR$ in Extended Data Fig. 1c.

The deciding factor that led to our best quality crystals was the use of higher purity W (99.999%) and Te (99.9999%) from Furuuchi Chemical Corporation. In addition the starting materials were loaded directly into the quartz tube, using a quartz wool filter instead of the frit-disc alumina crucibles. This allowed us to use larger amounts of starting materials, which should improve the volume to surface ratio of the melt in favor of a lower impurity density in the crystals from contact contamination with the quartz tube. Using our previously optimized mixing ratio of W:Te = 1:120, and a cooling rate of 2 °C/h from 1000 °C to 600 °C followed by centrifuging, we obtained crystals with excellent $RRR$ and $MR$ values on par with the best crystals reported in literature (see Extended Data Fig. 1c). We note that a cooling rate of 1 °C/h resulted in crystals with slightly lower quality.

## Device fabrication

WTe$_2$ crystals were mechanically exfoliated onto an oxidized silicon wafer (290 nm of SiO$_2$) and suitable flakes were identified by optical microscopy. Standard nano-fabrication techniques were used for device fabrication: electron beam lithography (EBL), inductively coupled plasma etching (ICP), and electron gun metal deposition (E-gun). Separate EBL steps were used to define the mesa and the contact geometries. The WTe$_2$ mesa was etched with ICP using SF$_6$ (15 sccm) and O$_2$ (5 sccm), RF power of 20 W, resulting in an etch rate of ~0.8 nm/s.

The Au reference samples and contacts to WTe$_2$ flakes were fabricated by E-gun deposition of Ti (2 nm) and Au (30 to 60 nm) followed by a lift-off procedure in acetone. Ar ion milling was used prior to the contact metal deposition for removal of the WTe$_2$ oxidation layer. For transport characterization, separate devices were fabricated in Hall bar geometry with width $W = 5$ μm, thickness $d \cong 40$ nm, and distance between the voltage contacts $L \cong 3.3$ μm.

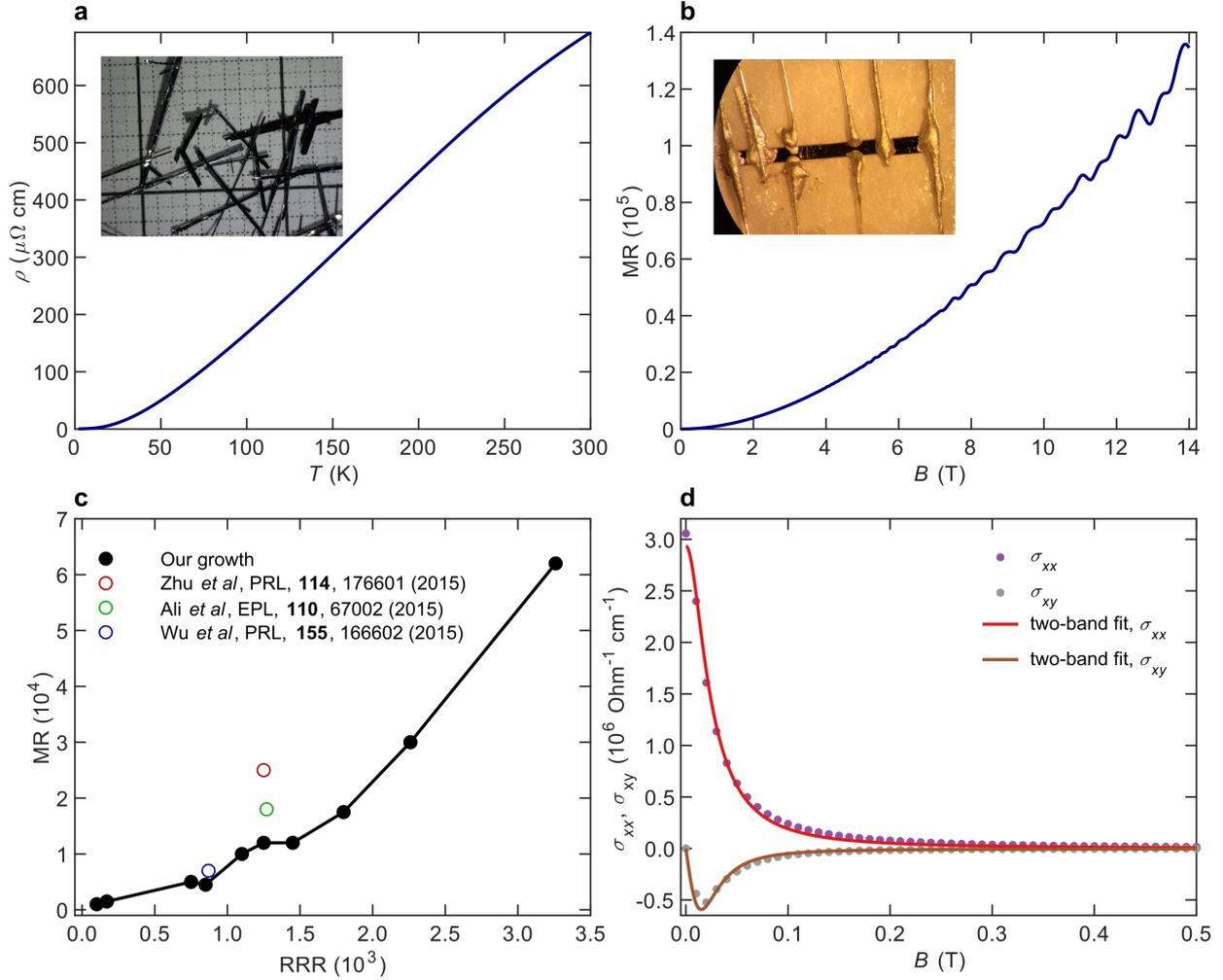

**Extended Data Fig. 1**. **Transport characterization of bulk WTe$_2$ single crystals. a**, Resistivity, $\rho$, as a function of temperature of our highest purity crystal. At $T = 2$ K, the resistivity is $\rho = 0.23$ μΩ·cm corresponding to $RRR \cong 3{,}250$. Inset: optical image of crystals from the optimized quality growth. **b**, Magnetoresistance, $MR = \frac{\rho(B) - \rho(0)}{\rho(0)}$, as a function of magnetic field at 2 K showing $MR \cong 62{,}000$ at 9 T. **c**, $MR$ vs. $RRR$ at $T = 2$ K and $B = 9$ T of our different crystals synthesized by flux growth (black dots) in comparison to reported values (open circles) in the literature [72–74]. The black line is a guide to the eye. **d,** Longitudinal and transverse conductivities $\sigma_{xx}$ and $\sigma_{xy}$ vs. magnetic field at 4.2 K and their fit to the two band model with resulting parameters $n_e = 2.4 \times 10^{19}$ cm$^{-3}$, $n_h = 2.3 \times 10^{19}$ cm$^{-3}$, $\mu_e = 5.1 \times 10^5$ cm$^2$/Vs, and $\mu_h = 2.7 \times 10^5$ cm$^2$/Vs.

**Magnetotransport measurements**

For bulk transport measurements, crystals with elongated geometry were selected with typical dimensions of width $W = 250$ to 350 μm, thickness $d = 22$ to 240 μm, and distance between the voltage contacts $L = 1$ to 3.7 mm. Electrical contacts were made with conductive silver epoxy resin (EPO-TEK H20E) using 50 μm diameter gold wire. The epoxy contacts were cured at 150 °C under continuous N$_2$ flow. An optical image of a representative crystal with current and voltage contacts is shown in the inset of Extended Data Fig. 1b. The transport measurements (temperature and field dependence of resistivity) were carried out in a physical property measurement system (PPMS, Quantum Design) using a *dc* current of 1 to 5 mA for bulk samples and an *ac* current of 100 nA at frequency of $f = 11.51$ Hz for crystal flakes. Transverse and longitudinal voltages were symmetrized and anti-symmetrized with respect to the magnetic field.

WTe$_2$ is a nearly compensated semimetal with electron and hole pockets contributing to transport [70,72–77]. We thus use a two-band conductivity model for the analysis of the magnetotransport:

$$\sigma_{xx}(B) = e\left(\frac{n_h \mu_h}{1+\mu_h^2 B^2} + \frac{n_e \mu_e}{1+\mu_e^2 B^2}\right), \quad \sigma_{xy}(B) = e\left(\frac{n_h \mu_h^2 B}{1+\mu_h^2 B^2} - \frac{n_e \mu_e^2 B}{1+\mu_e^2 B^2}\right),$$

where $e$ and $B$ are the elementary charge and applied magnetic field, and $n_e$, $n_h$, $\mu_e$, and $\mu_h$ are the electron and hole densities and mobilities, which are the fitting parameters.

Extended Data Fig. 1a shows the resistivity $\rho(T)$ measurement of a crystal from our best quality batch. The attained $RRR \cong 3{,}250$ slightly exceeds the values of $RRR = 900$ to 2500 reported in previous landmark studies [72–74,78]. Also, the magnetoresistance (MR) at 2 K (Extended Data Fig. 1b) shows an exceptionally high value of ~ 62,000 at 9 T, even exceeding the values of $MR \cong 42{,}000$ in WP$_2$ [79] and $MR \cong 17{,}500$ in WTe$_2$ [73], measured at the same field. At 14 T, our WTe$_2$ crystal attains $MR \cong 140{,}000$. In Extended Data Fig. 1c, we compare the $RRR$ and $MR$ values in some of our crystals from different batches with previously reported values. By fitting the conductivity at low fields to the two-band model, we obtained the electron and hole concentrations and their mobilities shown in Extended Data Fig. 1d. We took the average electron mobility in our bulk samples to be $\mu_e \cong 2.5 \times 10^5$ cm$^2$/Vs at 4.2 K, from which we derived the electron momentum-relaxing mean free path $l_{mr} = \hbar k_F \mu_e / e \cong 20$ μm ($k_F = 1.22$ nm$^{-1}$ is the Fermi wavelength [74]).

Transport measurements of our WTe$_2$ flakes in Hall bar geometry with thickness $d \cong 40$ nm show typical conductivities of $\sigma_{flake} \cong 8\times10^4$ Ohm$^{-1}$cm$^{-1}$ at 4.2 K, which are significantly lower than the bulk conductivities, $\sigma_{bulk} \cong 3\times10^6$ Ohm$^{-1}$cm$^{-1}$, leading to estimated effective $l_{mr} \cong 530$ nm in our flakes.

**SQUID-on-tip and magnetic imaging**

For the magnetic imaging measurements, Pb SOTs were fabricated with diameters ranging from 120 nm to 140 nm following the methods described in Ref. [13]. The SOTs were protected from oxidation by deposition of 3 to 5 nm thick Ti films below and on top of the Pb film. The SOTs included integrated shunt resistors on the tip [80] and had magnetic sensitivity of approximately 50 nT/Hz$^{1/2}$ in applied magnetic field of 60 mT. The SOT readout was carried out using a cryogenic SQUID series array amplifier (SSAA) [81–83]. For height control, the SOT was attached to a quartz tuning fork as described in Ref. [84].

Magnetic imaging was carried out at 4.5 K in 25 μbar residual He pressure in the chamber. For the measurements in Figs. 1 to 4, an *ac* current of $I_0 = 50$ μA at frequency $f = 186.4$ Hz was applied to the WTe$_2$ or Au samples and the corresponding out-of-plane component of the Oersted field $B_z(x, y)$ was measured by a lock-in amplifier at a constant height of 50 nm above the sample surface. For the scans in Fig. 5, the ac current was varied between $I_0 = 100$ μA and $I_0 = 400$ μA. The images were acquired with a pixel size of 13 nm, acquisition time of 40 ms/pixel, and image size of $430 \times 305$ pixels.

**Current density reconstruction**

For the reconstruction of the 2D current density $\mathbf{J}(x, y)$ from the measured $B_z(x, y)$, we have used the inversion method described in detail in Ref. [66]. The procedure allows for correction of a small possible tilt of the SOT from the vertical axis and for its finite size, and takes into account the finite thickness $d$ of the sample.

The inversion, however, is an ill posed problem and as such is prone to various artifacts, including high sensitivity to noise and fluctuations, boundaries of the imaging window, fields arising from sources outside the imaging window, ringing at sharp edges due to scanning height related low-pass filtering, and high sensitivity to the assumed height of the sensor. To stabilize the solution with respect to fluctuations, filtering and regularization methods are required [66,85]. As a result, the qualitative features of the resulting $J(x, y)$ are well reproduced; however, the precise quantitative details and the fine structure are less reliable. In the following, we detail the artifacts arising due to ringing and sensor height.

For controlling the scanning height, the SOT is attached to a quartz tuning fork (TF) [84]. The TF is exited electrically at its resonance frequency ~33 kHz and the shift in its phase is monitored as a function of height $h$ upon approaching the sample surface. A threshold of 1° phase shift is defined as the "poking" height, $h = 0$. We then retract the SOT and scan at a nominal height of $h = 50$ nm. We note that the actual effective height of the magnetic imaging should be larger due to the low phase shift threshold, possible surface residues, and the finite thickness of the Ti/Pb/Ti film of the SOT. In addition, the accuracy of the calibration of the vertical displacement of the piezoelectric scanner is limited.

Extended Data Fig. 2 demonstrates the effect of the assumed effective sensor height $h$ on the current distribution reconstructed from the measured $B_z(x, y)$ in the $\theta = 35°$ sample. It shows that the qualitative features of the current flow patterns, including the vortices in the chambers, are robust with respect to the assumed height in the range of $h = 20$ to 150 nm. Inspection of $J_y$ shows, as expected, that low $h$ of 20 and 50 nm results in some broadening of the derived current profiles (Extended Data Figs. 2g,h), whereas higher $h$ of 100 and 150 nm gives rise to enhanced ringing at the edges with negative current visible outside the sample edges (light blue in Extended Data Figs. 2i,j). This ringing is much less pronounced in the $J_x$ distribution in Extended Data Figs. 2b-d and is noticeable in Extended Data Fig. 2e predominantly near the chamber apertures. The ringing in $J_x$ is less significant because the value of $J_x$ in the strip and in the chambers is comparable and its absolute value is a much lower than $J_y$ in the strip. To minimize the ringing in $J_y$, we therefore use the nominal height $h = 50$ nm for the current reconstruction in all of the figures in the main text.

The ringing at the sharp edges, however, is an unavoidable feature of the low-pass filtering of the inversion procedure and is one of the limiting factors in determining quantitatively the accurate current profiles. This is exemplified by taking a uniform current density $J_y$ in an infinite strip of width $W = 550$ nm and thickness $d = 48$ nm (as expected in a Au strip). We then calculate numerically $B_z(x)$ at $h = 150$ nm, and perform numerical inversion back to current. The resulting reconstructed $J_y$ (green dots in Extended Data Fig. 3a) deviates substantially from the original uniform current density (light green line). It shows ringing both inside and outside of the strip and finite slope at the edges. The ringing artifacts depend on the various parameters of the inversion procedure, including the pixel size, but they cannot be eliminated. Thus deriving precise current profiles from the measured magnetic field has always a limited accuracy. The black solid line in Extended Data Fig. 3a shows the current profile reconstructed from the experimentally measured $B_z(x)$ across the Au strip assuming effective height of 150 nm. It shows a qualitative agreement with the green dotted curve, consistent with a uniform current distribution in the ohmic regime in Au. Assuming a lower effective height results in visible broadening of $J_y(x)$, while higher effective $h$ causes large oscillations.

In the strip geometry the current profile in the hydrodynamic regime is given by

$$J_y(x) = J_0 \left[ \frac{1 + (\xi/D)\tanh(W/2D) - \cosh(x/D)/\cosh(W/2D)}{1 + (\xi/D - 2D/W)\tanh(W/2D)} \right], \quad (2)$$

where $\xi$ is the slip length at the boundaries. For no slip conditions ($\xi = 0$), one obtains the familiar Poiseuille profile as shown by the light blue curve in Extended Data Fig. 3b. Even though the no-slip boundary conditions have been considered for the analysis of current profiles [7–10], it has been argued that they are not physical in electron fluids, and that $\xi$ of the order of $l_{ee}$ should be expected even in the case of fully diffuse boundaries [67]. From the analysis of vortex intensity in Fig. 3, we conclude $\xi = 200$ nm or larger in our samples. In this regime, the hydrodynamic current profiles in the strip geometry are almost indistinguishable from the ballistic profiles given by

$$J_y(x) = \frac{J_0}{\pi} \int_{-\pi}^{\pi} d\theta \cos^2\theta \left[ 1 - (1-r) \frac{\cosh(\csc\theta[W - 2x\,\mathrm{sgn}(\theta)]/2l_{mr}) + \sinh(\csc\theta[W - 2x\,\mathrm{sgn}(\theta)]/2l_{mr})}{-r + \cosh(W \cdot \csc|\theta|/l_{mr}) + \sinh(W \cdot \csc|\theta|/l_{mr})} \right], \quad (3)$$

where $0 < r < 1$ is the reflectivity coefficient, with $r = 0$ corresponding to fully diffuse boundaries and $r = 1$ describing specular boundaries. Extended Data Figs. 3e-g compare the ballistic profiles for $r = 0$, 0.5, and 1 (red) with the hydrodynamic case with $\xi = 200$ nm (blue). These results show that in a strip geometry, the difference between these cases based on the reconstructed $J_y(x)$ (blue and red dots) is small and experimentally insignificant. In contrast, in the chamber geometry, the vortex stability differs greatly between ballistic and hydrodynamic cases and is hardly affected by the boundary conditions, as shown in Fig. 3.

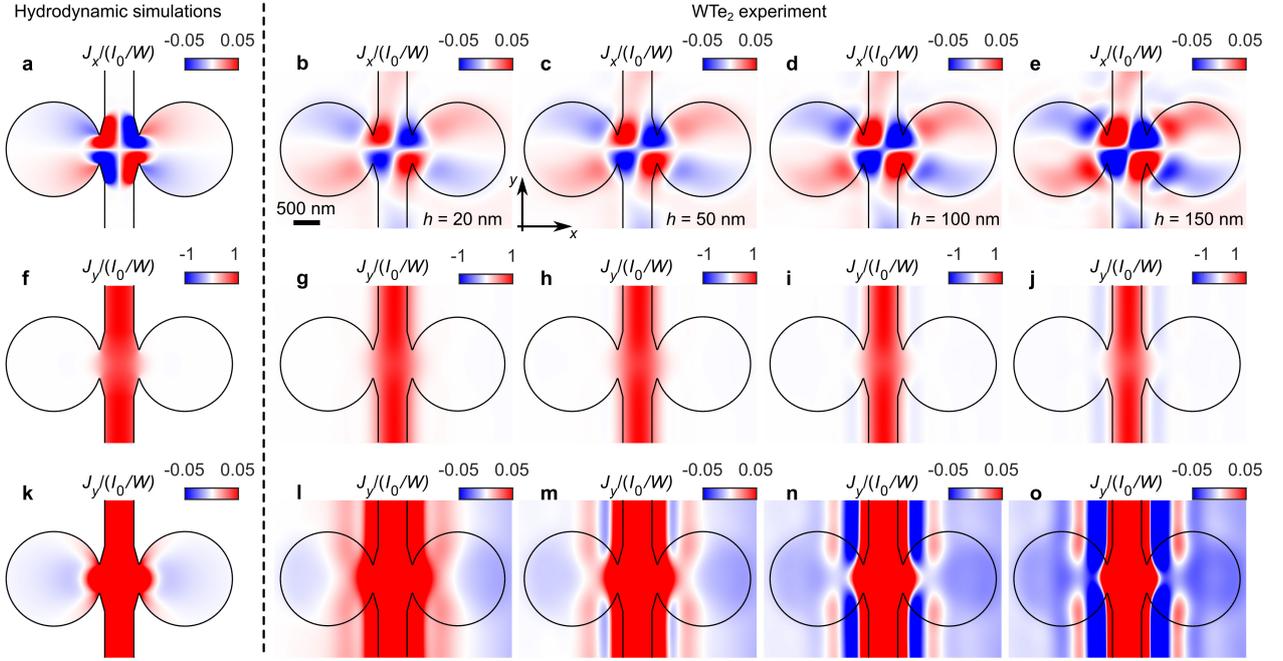

**Extended Data Fig. 2**. **Dependence of the reconstructed current densities on the assumed SOT scanning height.** **a**, Numerical simulation of $J_x(x,y)$ normalized by the average current density $I_0/W$ in the strip in $\theta = 35°$ sample for $D/W = 0.28$ and $\xi = 200$ nm. The span of the color scale is $\pm 0.05$. **b-e**, Current densities $J_x(x,y)$ reconstructed from the inversion of the measured $B_z(x,y)$ in WTe$_2$ sample A with $\theta = 35°$ assuming effective SOT scanning heights of $h = 20$ nm (**b**), 50 nm (**c**), 100 nm (**d**) and 150 nm (**e**). The nominal scanning height was 50 nm. The span of the color scale is $\pm 0.05$. **f-j**, Same as **a-e**, but for $J_y(x,y)$ on color scale of $\pm 1$. **k-o**, Same as **f-j**, but on expanded color scale of $\pm 0.05$. The $J_y$ vortex counterflow current (light blue) is resolved in the chambers on a large artificial ringing background outside the strip edges.

Taking the above limitations into account, we now inspect $J_y(x,y)$ more carefully within the chambers. Extended Data Fig. 2k shows the calculated $J_y$ with color scale expanded 20 times, such that the color of the laminar current in the strip (red) is strongly saturated. On this expanded scale, the $J_y$ counterflow becomes visible (light blue). Note that the density of the vortex current counterflow both in $J_x$ and $J_y$ is only about 1% of the laminar current density in the strip as seen in Extended Data Figs. 2a,k. By expanding the color scale of the experimental $J_y(x,y)$ in Extended Data Figs. 2l-o by the same factor of 20, the counterflow in $J_y$ (light blue) becomes visible. On this expanded scale the ringing in $J_y$ is very pronounced even for $h = 20$ nm and grows significantly with $h$, but the enhanced light blue signal of $J_y$ in the far side of the chambers is resolved at all values of $h$ and shows little dependence on $h$. Note that the experimental $J_x$ in Extended Data Figs. 2b-e is presented on the same color scale as $J_y$ in Extended Data Figs. 2l-o, demonstrating that the magnitude of the backflow in $J_y$ matches the scale of $J_x$ in the chambers, providing an independent confirmation for the observation of a vortex. Since the ringing problem in $J_x$ is much less pronounced, the counterflow of the vortex current is readily resolved in $J_x$ despite being only about 1% of the driving current density.

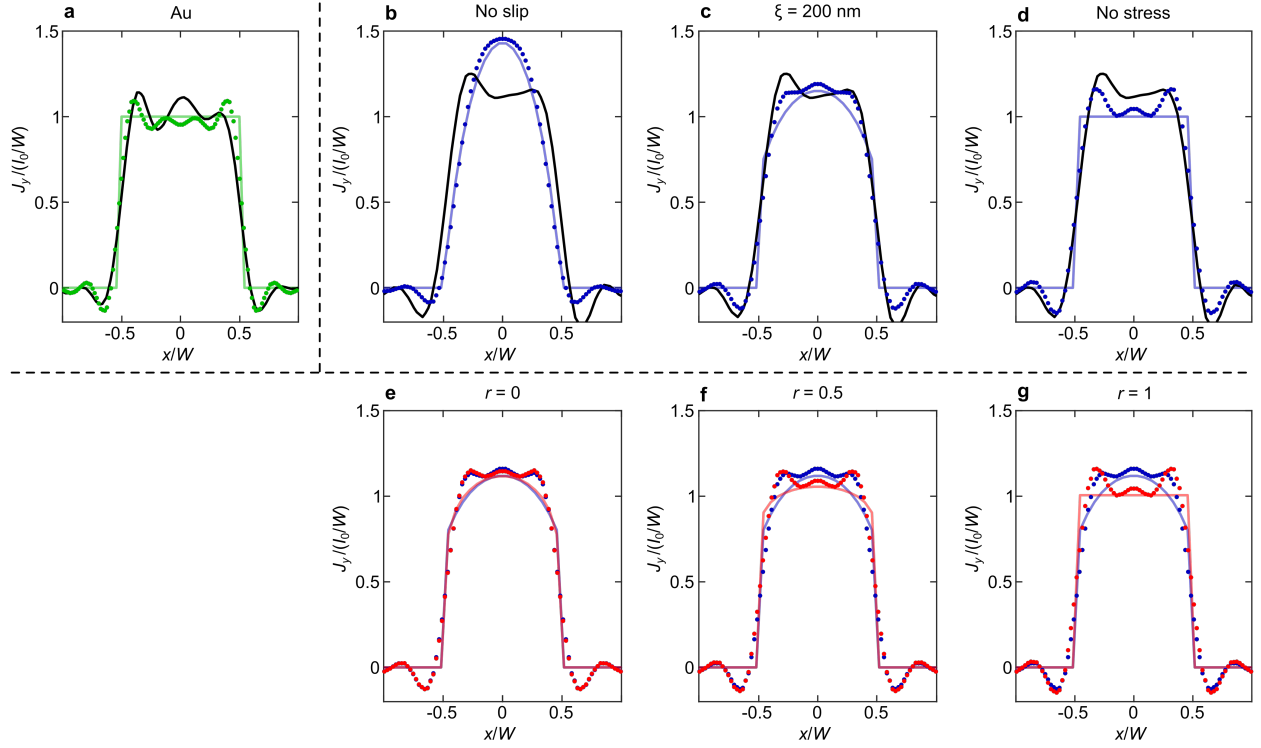

**Extended Data Fig. 3. Current profiles in narrow Au and WTe₂ strips. a**, A uniform current density $J_y(x)$ in $W = 550$ nm strip (light green line) from which $B_z(x)$ is computed at a height $h = 150$ nm. The $J_y(x)$ (green dotted symbols) is then reconstructed by inversion of the calculated $B_z(x)$, showing the unavoidable distortions and ringing. The $J_y(x)$ reconstructed from the experimental $B_z(x)$ in the Au strip (black line) shows consistency with a uniform current distribution in the ohmic regime. **b**, Same as (**a**) for a Poiseuille current profile (light blue) with $D/W = 0.28$ and no-slip boundary conditions. The reconstructed $J_y(x)$ from the experimentally measured $B_z(x)$ in WTe₂ strip (black) is inconsistent with the theoretically reconstructed $J_y(x)$ (dotted blue) from $B_z(x)$ corresponding to the Poiseuille profile. **c**, Same as (**b**) for hydrodynamic flow with $D/W = 0.28$ and slip length $\xi = 200$ nm (light blue) showing good correspondence between the theoretically reconstructed $J_y(x)$ (dotted blue) and the experimentally derived $J_y(x)$ (black) in accord with the conclusions in the main text. **d**, Same as (**b**) for hydrodynamic flow with $D/W = 0.28$ and no-stress boundary conditions (light blue). The reconstructed theoretical $J_y(x)$ (dotted blue) underestimates the experimentally derived $J_y(x)$ (black) supporting the conclusion of a finite slip length. **e**-**g**, Comparison between theoretically calculated current profiles in the hydrodynamic regime with $\xi = 200$ nm (light blue line) and in the ballistic flow (light red line) with boundary reflectivity coefficients of $r = 0$ (fully diffuse) (**e**), $r = 0.5$ (**f**), and $r = 1$ (specular) (**g**). The solid lines show $J_y(x)$ calculated from Eqs. 2 and 3 while the dotted lines are the current profiles reconstructed from the calculated corresponding $B_z(x)$. These results demonstrate the difficulty in using reconstructed current profiles in strip geometry for distinguishing between the hydrodynamic flow with finite slip length and the ballistic transport, in contrast to vastly different vortex stability phase diagrams in these two regimes.

**COMSOL numerical simulations**

The 2D finite-element numerical simulation of an ohmic electron flow and of transport described by Eq. 1 for the ohmic, ballistic and hydrodynamic regimes, as discussed in the main text, were carried out using COMSOL Multiphysics 5.4. We used the Coefficients Form PDE module, which solves the general equation:

$$e_a \frac{\partial^2 \boldsymbol{u}}{\partial t^2} + d_a \frac{\partial \boldsymbol{u}}{\partial t} + \nabla \cdot (-c \nabla \boldsymbol{u} - \alpha \boldsymbol{u} + \gamma) + \beta \cdot \nabla \boldsymbol{u} + a \boldsymbol{u} = f,$$

where the field $\boldsymbol{u}$ is:

$$\boldsymbol{u} = \begin{pmatrix} \phi \\ J_x \\ J_y \end{pmatrix},$$

and the coefficients were chosen to match Eq. 1:

$$e_a = d_a = \alpha = \gamma = f = 0,$$

$$c = \begin{pmatrix} 0 & 0 & 0 \\ 0 & D^2 & 0 \\ 0 & 0 & D^2 \end{pmatrix}, \quad a = \begin{pmatrix} 0 & 0 & 0 \\ 0 & 1 & 0 \\ 0 & 0 & 1 \end{pmatrix}, \quad \beta = \begin{pmatrix} \begin{pmatrix} 0 \\ 0 \end{pmatrix} & \begin{pmatrix} 1 \\ 0 \end{pmatrix} & \begin{pmatrix} 0 \\ 1 \end{pmatrix} \\ \begin{pmatrix} \sigma \\ 0 \end{pmatrix} & \begin{pmatrix} 0 \\ 0 \end{pmatrix} & \begin{pmatrix} 0 \\ 0 \end{pmatrix} \\ \begin{pmatrix} 0 \\ \sigma \end{pmatrix} & \begin{pmatrix} 0 \\ 0 \end{pmatrix} & \begin{pmatrix} 0 \\ 0 \end{pmatrix} \end{pmatrix},$$

where $D$ is the Gurzhi length, $\sigma$ is the conductivity, and the source and drain are simulated by Dirichlet boundary conditions on the potential $\phi$. The resulting current density field $\boldsymbol{J}(x,y)$ was normalized by the average current density $J_0 = \int_W J_y(x,y) dx / W$ in the strip. A complete description of the problem is given by employing boundary conditions. The boundary conditions for the perpendicular current were $\boldsymbol{J}_\perp = \boldsymbol{J} \cdot \hat{n} = 0$ in all simulations. For the tangential current, $\boldsymbol{J}_\parallel = (\boldsymbol{J} - \boldsymbol{J}_\perp \cdot \hat{n})$, we have used the three types of boundary conditions described in the main text: (*i*) no slip, $\boldsymbol{J}_\parallel = 0$, (*ii*) no stress, $\hat{n} \cdot \nabla \boldsymbol{J}_\parallel = 0$, and (*iii*) a finite slip length, $\boldsymbol{J}_\parallel = \xi \hat{n} \cdot \nabla \boldsymbol{J}_\parallel$. The dimensions of the simulated devices were chosen to be equal to those of our samples, with the width of the central strip $W = 550$ nm in the dual chamber geometry and the disk radius of the chambers of $R = 900$ nm.

For derivation of the vortex stability phase diagram, a geometry depicted in Fig. 3 was used, with variable $D$ and $\theta$. The vortex current $I_v$ for each simulated geometry and $D$ was calculated according to

$$I_v(D, \theta) = \frac{1}{4} \int_l \left( |J_y(x,0)| - J_0 \right) dx,$$

where the integral was carried out along the cross-section through the central horizontal line connecting the centers of the disk chambers. For $D/W > 1$, the maximum of $I_v$ occurs at $\theta = 60°$ when the aperture and the radius of the disk chambers form an equilateral triangle. An equilateral triangle in the context of hydrodynamic flow was also reported in Ref. [23]. Upon $D/W$ decreasing, the maximum point of $I_v$ shifts to lower $\theta$.

Supplementary Videos 1 and 2 show the evolution of the streamlines upon decreasing $\theta$ in the hydrodynamic and ballistic regimes, respectively. As $\theta$ increases, on approaching the vortex-to-laminar transition, the laminar streamlines (red) penetrate deeper into the chambers causing distortion of the vortex into a banana shape. In the ballistic case the vortex (blue streamlines) is pushed out of the chamber as a whole, while in the hydrodynamic regime it splits into two vortices at the top and bottom of the chambers. This enhanced stability of the two-vortex solution in the hydrodynamic regime is well captured by the analytical estimates presented in the theory section below.

**Theory**

*Angular scattering, diffusion along the Fermi surface and para-hydrodynamics*

This section aims to provide a microscopic justification of Eq. 1 which is used as a benchmark model in the main text. We consider a Fermi gas in two dimensions with a cylindrically symmetric dispersion and a circular Fermi surface. Transport in the system weakly perturbed away from equilibrium is described by a steady-state carrier distribution of the form $\bar{f}(p,r) = \bar{f} + \delta f(r, \varphi)$, with $\bar{f}$ being the equilibrium distribution and $\delta f(r, \varphi)$ a perturbation, where $\varphi$ is the angle on the Fermi surface, $p$ is electron momentum and $r = (x, y)$. The steady-state distribution satisfies the linearized kinetic equation:

$$v_F \cos \varphi \frac{\partial \delta f}{\partial x} + v_F \sin \varphi \frac{\partial \delta f}{\partial y} - \hat{I} \delta f = -\mathbf{F} \cdot \nabla_p \bar{f} \qquad (4)$$

Here $\hat{I}$ is the linearized collision operator of the elastic scattering and **F** is an external force. Equation 4 is valid for a collision operator of a general form; below, we apply it to describe scattering at the sample upper and lower surfaces, the process discussed in the main text. Following [59], we expand our perturbation over angular harmonics:

$$\delta f(y, \varphi) = \int \frac{d^2 k}{(2\pi)^2} \sum_{m=-\infty}^{m=\infty} f_m(k) e^{im\varphi + i\mathbf{k}\mathbf{r}}.$$

Statistical isotropy of scattering means that the angular harmonics are eigenfunctions of the scattering operator: $\hat{I} f_m = -\gamma_m f_m$. Because of the $\cos\varphi$ and $\sin\varphi$ structure of the streaming term in Eq. 4, the harmonics $f_m(k)$ satisfy a tridiagonal system of linear equations. The contribution generated by the $m$-th harmonic of the right-hand side of Eq. 4 (denoted $B_m$) satisfies the equation:

$$\gamma_m f_m + iA f_{m+1} + i\bar{A} f_{m+1} = B_m, \tag{5}$$

where $A = (k_x - ik_y)v_F/2$. Following [59], we solve Eq. 5 for the ratios $\alpha_m = if_{m+1}/f_m$, which for $n \neq m$ satisfy the recursive relation

$$\alpha_{n-1} = \frac{\bar{A}}{\gamma_n + A\alpha_{n-1}}.$$

The solution of this problem is given in a closed form as a continued fraction

$$\alpha_{n-1} = \cfrac{\bar{A}}{\gamma_n + \cfrac{|A|^2}{\gamma_{n+1} + \cfrac{|A|^2}{\gamma_{n+2} + \cdots}}}$$

Similarly for $\beta_m = if_{m-1}/f_m$ we obtain the fraction running down:

$$\beta_{n+1} = \cfrac{\bar{A}}{\gamma_n + \cfrac{|A|^2}{\gamma_{n-1} + \cfrac{|A|^2}{\gamma_{n-2} + \cdots}}}.$$

Substituting into Eq. 5 yields a contribution to the $m$-th harmonic of perturbed distribution as follows:

$$f_m = \cfrac{B_m}{\gamma_m + \cfrac{(kv_F/2)^2}{\gamma_{m-1} + \cfrac{(kv_F/2)^2}{\gamma_{m-2} + \cdots}} + \cfrac{(kv_F/2)^2}{\gamma_{m+1} + \cfrac{(kv_F/2)^2}{\gamma_{m+2} + \cdots}}}. \tag{6}$$

Considering long-wavelength limit, we retain only the terms up to quadratic in wavenumber, which gives dissipation and diffusion terms:

$$\left(\gamma_m + \frac{(kv_F)^2}{4\gamma_{m-1}} + \frac{(kv_F)^2}{4\gamma_{m+1}}\right) f_m = B_m.$$

For $m = 0$, one usually has $\gamma_0 = 0$ due to charge conservation, so that we have pure diffusion. For a locally homogeneous electric field we have

$$v_F \cos\varphi \frac{\partial \delta f}{\partial y} + v_F \sin\varphi \frac{\partial \delta f}{\partial x} - \hat{I}\delta f = -e\frac{\partial \bar{f}}{\partial \epsilon}\left(v_F \cos\varphi \frac{\partial \phi}{\partial y} + v_F \sin\varphi \frac{\partial \phi}{\partial x}\right)$$

$$\gamma_m f_m + iA f_{m+1} + i\bar{A} f_{m+1} = B\delta_{m,1} + \bar{B}\delta_{m,-1}.$$

Here, $2B = e(E_x + iE_y)v_F \partial \bar{f}/\partial \epsilon$, where $\epsilon$ is electron energy. In this case, Eq. 6 is the current-field relation with a nonlocal conductivity:

$$\mathbf{J}_k = \sigma(k)\mathbf{E}_k, \tag{7}$$

$$\sigma(k) = \frac{ne^2/m}{\gamma_1 + \Gamma(k)}, \qquad \Gamma(k) = \frac{(kv_F/2)^2}{\gamma_2 + \frac{(kv_F/2)^2}{\gamma_3 + \cdots}}. \tag{8}$$

In the long-wavelength limit set by the $m=3$ harmonic decay rate, such that $(kv_F)^2 < 4\gamma_2\gamma_3$, Eqs. 7 and 8 give

$$\frac{(kv_F)^2}{4\gamma_1\gamma_2} \boldsymbol{J_k} + \boldsymbol{J_k} = -\sigma \boldsymbol{E_k}. \tag{9}$$

After applying a Fourier transform, we recover Eq. 1 from the main text with $D = v_F/\sqrt{4\gamma_1\gamma_2}$:

$$-D^2\nabla^2 \boldsymbol{J} + \boldsymbol{J} = -\sigma\nabla\phi. \tag{10}$$

Notably, this equation is applicable not only in the hydrodynamic and ohmic regimes but also in the ballistic regime. This behavior is unique to the situation when $\gamma_3 \gg \gamma_2$, since in this case the condition for the length scales $(kv_F)^2 < 4\gamma_2\gamma_3$ used to derive Eq. 10 is valid in both the hydrodynamic and ballistic regimes. This is in contrast to the conventional electron fluids, where Eq. 10 is valid only in the hydrodynamic and ohmic regimes, but not in the ballistic regime. The extended validity range of Eq. 10 is a salient feature due to small-angle scattering.

The relation between our interpretation of the observed hydrodynamic behavior in terms of Eq. 10 and the assumption that the rates $\gamma_1, \gamma_2, \gamma_3 \ldots$ are determined by the small-angle scattering on sample surfaces can be further substantiated as follows. The requirement for Eq. 10 to hold is $\gamma_3 \gg \gamma_2 \gg \gamma_1$. This condition can be approximately fulfilled for small-angle scattering that leads to angular diffusion. Indeed, in this case we have $\hat{I} \approx \gamma \frac{\partial^2}{\partial\varphi^2}$, and therefore $\gamma_m = \gamma_1 m^2$. The condition for the long-wavelength limit is then $kv_F < \sqrt{4\gamma_2\gamma_3} = 12\gamma_1$. In terms of the effective momentum-relaxation length $l_{mr} = v_F/\gamma_1$, the condition takes the form $kl_{mr} < 12$, which is not too restrictive. Indeed, if we put $k \cong 1/W$, the Eq. 10 with $D = l_{mr}/4$ is expected to work reasonably well for $l_{mr} < 12W$, a regime well satisfied for our parameters. In contrast, for large-angle scattering, the decay rates for different harmonics take similar values, $\gamma_3 \cong \gamma_2 \cong \gamma_1$. In this case, depending on the size of $(kv_F)^2$, Eq. 8 can only be truncated at zeroth order or never, thus precluding a regime where Eq. 10 holds.

*Hydrodynamic vs. ballistic vortex formation: general considerations and scaling analysis*

The discussion in this section provides qualitative arguments in support of the hydrodynamic origin of the observed vortices, as suggested by our simulation results. In a system of size $W$, the flow pattern depends on two dimensionless parameters, $l_{ee}/W$ and $l_{mr}/W$. The hydrodynamic regime occurs when $l_{ee}/W \to 0$ and $l_{mr}/W \to \infty$, whereas the ballistic regime corresponds to $l_{ee}/W \to \infty$ and $l_{mr}/W \to \infty$, and the ohmic regime takes place when both $D/W \to 0$ and $l_{mr}/W \to 0$. To gain insight into the character of the flow patterns in all three limits we consider the dissipation in the flow: $\int dxdy [\sigma|J|^2 + \sum_{ij}(\sigma_{ij})^2/2\eta]$. Here, the first term is the ohmic dissipation, while the second term is the viscous dissipation, determined by the stress $\sigma_{ij} = \eta(\partial v_i/\partial x_j + \partial v_j/\partial x_i)$. The physical flow can be obtained by minimizing this dissipation functional supplemented with suitable boundary conditions, a procedure known as the principle of minimum entropy production.

The flow in the ohmic limit tends to minimize the current density everywhere, while the ballistic limit tends to minimize stress globally. Finally, the hydrodynamic limit minimizes stress locally, but not globally (Extended Data Fig. 4). As discussed below, the vortex stability phase boundary and the crossovers between different regimes can be obtained directly from the analysis based on the principle of minimum entropy production.

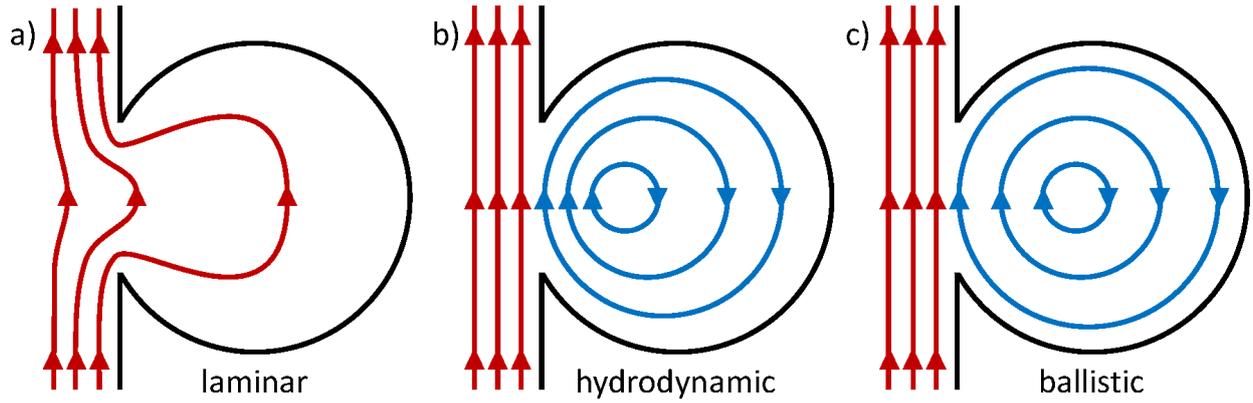

**Extended Data Fig. 4. Schematic streamlines for purely ohmic, hydrodynamic and ballistic flow. a**, If the sample is purely ohmic, the current leaks into the chamber, forming a current dipole decaying as inverse distance squared. **b**, For a purely hydrodynamic flow, no laminar current (red) leaks from the strip into the chamber. Instead, a vortex forms in the vicinity of the aperture in the chamber (blue) in order to decrease the shear due to the gradient in the velocity profile. **c**, In a purely ballistic flow, only the geometry dictates the streamlines, producing a vortex (blue) whose center is positioned near the chamber center.

*Hydrodynamic-to-ohmic and ballistic-to-ohmic crossovers*

Since there is no vortex in the ohmic regime, the transition to this regime (upon a decrease in $l_{mr}$, for instance) from either hydrodynamic or ballistic flow can be understood as a weakening of the vortex in the chamber. As vortex weakens, the laminar current $j_l$ from the strip is expected to penetrate deeper and deeper into the chamber. As a result, the laminar streamlines superimpose onto the vortex streamlines in panels (b) or (c) in Extended Data Fig. 4. In the core of the ballistic or hydrodynamic vortex, the superimposed laminar flow points in the direction of the strip flow, which means that the combined flow pattern in the presence of a small laminar component has its vortex center (where the current vanishes) displaced further away from the aperture compared to the purely ballistic or hydrodynamic case.

In the following analysis, we consider three components that comprise the total current density: laminar current $j_l$, hydrodynamic current $j_{hyd}$, and ballistic current $j_{bal}$. The aperture presents a small dipolar source and sink of the laminar component of the current, which leaks from the strip into in the chamber. Along the chamber boundary and as a function of the distance $\delta$ from the aperture, the density of the laminar component thus decays as $j_l \sim 1/\delta^2$ (for $l_{mr} > W$). In contrast, the vortical components (blue in Extended Data Fig. 4) do not flow into or out of the strip, but circulate entirely within the chamber. The ballistic vortex is almost rotationally symmetric and is positioned close to the center of the chamber. As a result, its density along the chamber boundaries is almost constant, $j_{bal} \sim const$. The hydrodynamic vortex, in contrast, forms directly outside the aperture, producing a flow that decays as $j_{hyd} \sim 1/\delta$ far outside the vortex core. Thus, if the laminar flow is strong enough to push the vortex core out to the far side of the chamber, it inevitably overpowers the flow in the upper and lower segments of the chamber, meaning that no vortical flow remains in either the hydrodynamic or the ballistic case. However, for hydrodynamic flow, the streamlines are not uniquely fixed by the geometry, and more than one vortex might form. This is indeed the case, as we demonstrate next: in the hydrodynamic regime, vortical flow can be stabilized by creating two vortex cores even if the single, large vortex is annihilated. In contrast, ballistic flow cannot form two vortex cores next to each other, because ballistic trajectories intersect each other without any effect, which means that the vortex cores do not repel each other, and instead merge into a single large vortex.

*Principal components of hydrodynamic flow*

To illustrate the basic physics behind the appearance and disappearance of vortices in the chamber, we develop a simple model which allows us to estimate dissipation and choose a state that minimizes it. We want to model the vortex that forms when current in a strip of width $W$ leaks into a large chamber or open space through an aperture in the strip wall at $x = 0$ of size $\Delta = 2R\sin(\theta/2)$, where $R$ is the chamber radius. The relevant current densities are $j_c$ in the strip, $j_v$ of the vortex, and $j_l$ the laminar current in the open space. These current densities are taken to be additive, where the relative amplitudes are determined by the requirement that the resulting flow minimizes the dissipation.

First we consider laminar flow, which can be modelled as a dipole describing current that flows out into the $x > 0$ half-plane in the interval $-\frac{\Delta}{2} < y < 0$ and returns in the interval $0 < y < \Delta/2$. For concreteness, we take the profile of the outflow through the aperture to be parabolic of the form $\sim y(|y| - \Delta/2)$, with the $y$ dependence sign-changing with a zero net current. For weak ohmic dissipation, a laminar ballistic flow in a half plane injected through an aperture with such a profile yields

$$\boldsymbol{j}_l(x, y) = j_{l0} \int_{-\Delta/2}^{\Delta/2} dy' \frac{y'(|y'| - \Delta/2)}{(y - y')^2 + x^2} \begin{pmatrix} x \\ y - y' \end{pmatrix}, \quad (11)$$

with the $x$ component odd in $y$ and the $y$ component even in $y$. This is a dipole source, and the integral can be done analytically. Expanding in large $x$ for $y = 0$ yields $|\boldsymbol{j}_l(x, 0)| = \frac{j_{l0}\Delta^4}{6x^2}$, meaning that the current decays inversely with distance squared at large distances from the aperture, as expected.

Next we consider a vortex positioned inside the chamber in proximity to the aperture. We assume that it has the shape of a Kaufmann vortex, with current density in polar coordinates given by $\boldsymbol{j}_v(r, \varphi) = j_{v0} rR_c/(r^2 + R_c^2)\hat{\boldsymbol{e}}_\varphi$. Here, $R_c$ is the size of the vortex core, and $j_{l0}$ and $j_{v0}$ measure the strength of the laminar and vortical flows, respectively. We assume that the vortex is located at $x = d_v$, and account for the boundary conditions close to the walls by adding a counter-rotating image vortex centered around $x = -d_v$. We further impose a condition that the total current $I_0$ is conserved in any cross-section with a fixed $y$, which leads to the condition that in the presence of the laminar outflow into the half-space, the current in the strip at $y = 0$ becomes $I_0 - c_{c\rho}j_l\Delta$, where $c_{c\rho}$ is a dimensionless geometrical factor.

Turning to the discussion of dissipation, we note that the decreased current density in the strip also reduces viscous dissipation, which we account for by subtracting the current densities from both the laminar current and the vortical flow. The current profile across the strip is parabolic if $D \sim W$. However, in the hydrodynamic limit when $D \ll W$, the current profile becomes very flat and the viscous dissipation is only relevant in a small boundary region. In this latter case, the term for the viscous dissipation acquires an additional factor of $D$ in the denominator. We therefore write an approximate expression for the total dissipation in our system area which is affected by the chamber (i.e. for the chamber itself and for the strip section between $-\Delta/2$ and $\Delta/2$), in which we treat the laminar and vortical flows as additive contributions. This gives

$$P(\boldsymbol{j}_l, \boldsymbol{j}_v, R_c) = P_{\rho,W} + P_{v,W} + P_{\rho,R} + P_{v,R} + P_{\rho,mixed} + P_{v,mixed}. \quad (12)$$

Here, ohmic dissipation $P_\rho = P_{\rho,W} + P_{\rho,R} + P_{\rho,mixed}$ measures the local current density squared, integrated over the system area, while the viscous dissipation $P_v = P_{v,W} + P_{v,R} + P_{v,mixed}$ measures the square of the current density gradients. For the central strip, the effective area is $W\Delta$ for current in the strip, while it is $\Delta^2$ for the laminar flow that penetrates from the strip into the chamber, and $R_c^2$ for the vortical flow. For the viscous dissipation, we note that the gradients are usually smooth and essentially cancel the area integral. Only in the strip, for small $D \ll W$, this is not true and the current density gradients are restricted to a narrow boundary region of size $D$. Therefore, $P_{\rho,W} = \frac{\rho}{2}(I_0 - c_{c\rho}j_l\Delta)^2\frac{\Delta}{W}$ is the ohmic dissipation in the strip section, while $P_{v,W} = \frac{\eta}{2}(\frac{I_0}{W} - c_{c\eta}(j_l + j_v))^2\frac{\Delta}{\min[D,W]}$ is the viscous dissipation in the same area ($\rho$ is the sheet resistance). The leading term, proportional to $\frac{\rho}{2}I_0^2$, does not affect the competition between the vortex and no-vortex states. Analogously, in the chamber we find $P_{\rho,R_c} =$ $\frac{\rho}{2}(c_{l\rho}j_l^2\Delta^2 + c_{v\rho}j_v^2 R_c^2)$ and $P_{v,R_c} = \frac{\eta}{2}(c_{l\eta}j_l^2 + c_{v\eta}j_v^2)$, up to logarithmic corrections in $R/W$. Here, $c_{c\eta}$, $c_{c\eta}$, $c_{l\rho}$, $c_{v\rho}$, $c_{l\eta}$, $c_{v\eta}$ are dimensionless form factors.

One implication of the construction of the flow in terms of $j_l$ and $j_v$ is that for both ohmic and viscous dissipation, the mixed terms in Eq. 12 between both components are negligible, a property that greatly simplifies the analysis. From our general considerations, we know that the effective size of the vortex goes to zero as the chamber opening is made smaller, i.e. $R_c(\Delta \to 0) = 0$. As the first main observation, we thus find that $P_{\rho,R_c}(\Delta \to 0) = 0$. In contrast, the viscous dissipation $P_{v,R}$ does not depend explicitly on the size of $\Delta$ (and $R_c$), and therefore the ratio $P_{\rho,R_c}/P_{v,R_c} \to 0$, i.e. for small apertures, the ohmic dissipation is negligible compared to the viscous one. Consequently, for small opening sizes $\Delta$, the solution is independent of $P_{\rho,R_c}$. This suffices to find a unique solution to $j_{l0}$ and $j_{v0}$ (cf. Eq. 11 and below), and subsequently construct an upper bound for the existence of a single vortex depending on whether the total dissipation is smaller with rather than without vortical flow. The vortex is energetically favorable as long as

$$\frac{\Delta_1}{W} < \frac{c_{l\eta}c_{v\eta}}{c_{c\eta}^2(c_{l\eta} + 2c_{v\eta})}\frac{\min[D,W]}{W}. \tag{13}$$

This result provides the general form of the vortex stability phase diagram: for $D/W \ll 1$, the vortex to no-vortex phase transition line is linear with $\Delta_1/W \cong \theta$, while for $D/W > 1$, $\theta$ saturates at a finite value, consistent with the numerically derived phase diagram in Figs. 3a,b.

Using the same ansatz, but starting from two vortex cores, the same condition Eq. 13 is recovered for the limiting opening size $\Delta_2$, but where $c_{v\eta}$ is replaced by $4c_{v\eta}$, (the factor 4 is due to the square of the current which enters in the dissipation) to account for the viscous dissipation from two cores. Since it holds that $\Delta_2 > \Delta_1$ this means that there is a narrow range of opening sizes where not one but instead two vortices can form. We note that by the same argument, even more vortex cores might be stabilized. However, at the same time, the positive effect of the vortex formation on the stress at the strip opening diminishes, making these more exotic solutions incompatible with the geometrical constraints imposed by the chamber. Numerically, all form factors are of order 1 and depend little on the particular choices of integration cutoffs. Specifically, for the dipole laminar flow and a Kaufmann vortex, we find that $c_{l\eta} \approx 4.0$ and $c_{v\eta} \approx 2.0 + 0.8\log(W/R_c)$. We further estimate that $W/R_c \sim 3$ and $c_{c\eta} \sim 1/2$, in which case $\Delta_1 < 4.7\min[D,W]$ and $\Delta_2 < 6.8\min[D,W]$. These numbers are reasonably close to both the numerical and experimental findings, even though the ratio of $\Delta_2/\Delta_1$ seems to be somewhat smaller in the simulations. In summary, for weakly ohmic flow, a single hydrodynamic vortex can form inside the chamber close to the aperture, with a size of vortex core that increases with the aperture size. For large apertures, this vortex becomes unstable, and we find a narrow range of parameters where it is favorable to form two vortices instead. Vortical flow disappears once the aperture size becomes sufficiently large to allow spreading of the laminal currents over the entire area of the chamber.

*Electron-electron scattering length in a compensated semimetal*

Here, we estimate the electron-electron scattering rate using the fermionic self-energy calculated in the random phase approximation (RPA). To that end, we consider three contributions that might be relevant in reducing $l_{ee}$ below the values reported in [10]. They are all related to the compensated nature of WTe$_2$, where the band edges of several bands are close to (but not at) the Fermi level.

(i) Firstly, the proximity of the band edges violates the core requirement of semiclassical estimates of the relaxation rate that the ratio of temperature over Fermi energy, $T/E_F \ll 1$, so that a fully quantum-mechanical treatment is needed. In this latter approach [86], the relaxation rate explicitly contains the occupation functions which account for both virtual and thermal fluctuations of the electron fluid. Given a dispersion $\epsilon_{mk}$ and eigenfunctions $|u_{mk}\rangle$, where $m$ is the band index and $k$ the momentum, the imaginary part of the self-energy $\Sigma$ assumes the form,

$$\text{Im}\,\Sigma(q,\epsilon_{nq}) = \sum_m \int \frac{d^3k}{(2\pi)^3}\,\text{Im}\left[\frac{C(k-q)|\langle u_{nq}|u_{mk}\rangle|^2}{1-\Pi(k-q,\epsilon_{mk}-\epsilon_{nq})C(k-q)}\right]\bigl(b(\epsilon_{mk}-\epsilon_{nq})+n(\epsilon_{mk})\bigr). \tag{14}$$

Here $b$ and $n$ denote the Bose and Fermi functions, respectively. The dielectric function is determined by the charge susceptibility $\Pi$ in the RPA-approximation,

$$\Pi(q,\omega) = \sum_{ij}\int \frac{d^3k}{(2\pi)^3}|\langle u_{ik}|u_{j\,k+q}\rangle|^2\,\frac{n(\epsilon_{ik})-n(\epsilon_{j\,k+q})}{i0^+ + \omega + \epsilon_{ik}-\epsilon_{jk+q}}, \tag{15}$$

while the Coulomb interaction is

$$C(k) = \frac{4\pi e^2}{|k|^2}. \tag{16}$$

(ii) Secondly, the presence of band edges also precludes the extrapolation of relaxation rates obtained at high temperatures under the assumption of a simple Fermi-liquid $T^2$-scaling, at least *a priori*. We find in particular that the $T^2$-dependence is violated for temperatures above 100 K.

(iii) Thirdly, the compensated nature of the material with both hole and electron Fermi surfaces requires a high-fidelity calculation with a fine momentum space grid to properly resolve the nesting between electron and hole Fermi surfaces. For example, we observe convergence of the obtained relaxation rates only for grid sizes larger than 100×50×7 in the $x$-$y$-$z$ momentum volume.

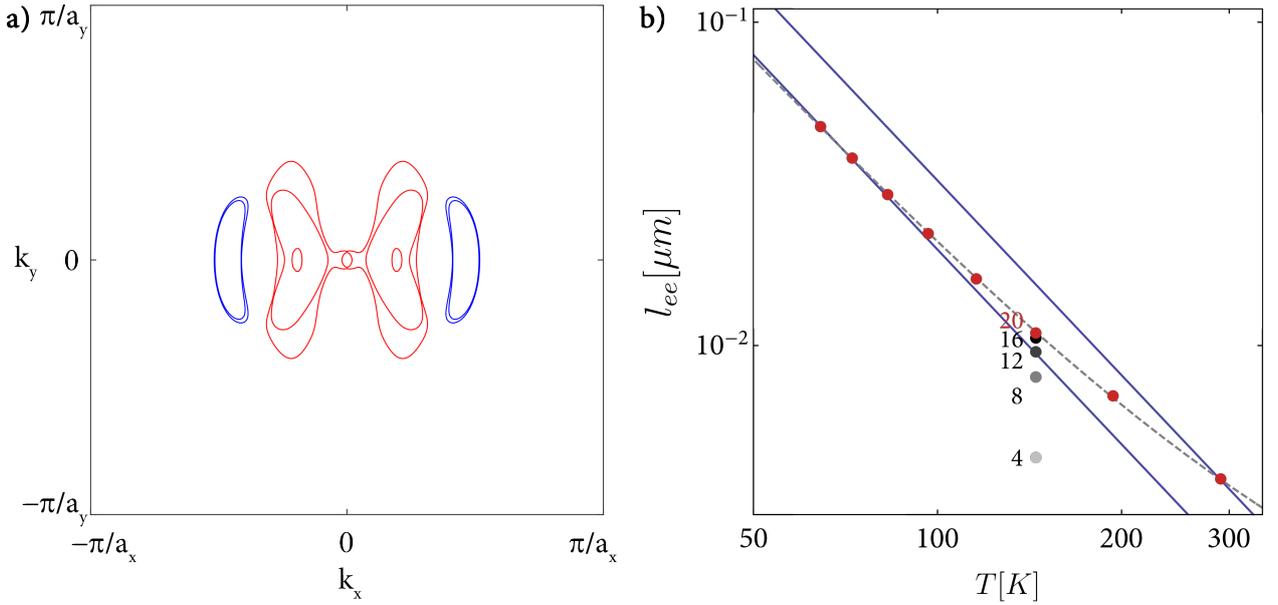

**Extended Data Fig. 5. Fermi surface and electron-electron mean free path. a**, Fermi surface cut for $k_z = 0$. Typical for a compensated semimetal, small electron and hole pockets appear close to the compensation point. If the hole density is slightly larger than the electron density, the Fermi surface features hole pockets near the Gamma point (red) and electron pockets (blue). **b**, $l_{ee}$ as calculated from Eq. 14 for 20 bands as a function of temperature (red points). For $T = 145$ K, we also show the values for a smaller number of bands. The blue lines denote upper and lower estimates for the $T^{-2}$ dependence of $l_{ee}$, where the lower one corresponds to the low-temperature asymptotics.

Taking all these issues into account, we calculated $l_{ee}$ for temperatures between 70 K and 300 K, checked for convergence in terms of grid resolution, and extrapolated to lower temperatures based on a power-law fit. To decrease runtime, as an approximation we restricted the effects of the Coulomb interaction to the first Brillouin zone and set the imaginary part of the dielectric function to zero for bands far from the chemical potential. Extended Data Fig. 5 shows the Fermi surface near the compensation point, and the scaling of $l_{ee}$ with temperature, averaged for band 55 which forms one of the hole pockets, taking into account a total of 20 bands closest to the chemical potential. Comparable values are obtained for the other Fermi pockets. As is clearly visible, the low-temperature asymptotic temperature dependence sets in only below 100 K. We also confirmed the convergence of our calculation with respect to contributions from far bands by comparing the obtained relaxation rates for 4, 8, 12, 16 and 20 bands at an intermediate temperature of 145 K. Extrapolating our results to low temperatures under the assumption of a $T^{-2}$ dependence, we obtain $l_{ee} = 0.5$ µm at $T = 20$ K and $l_{ee} = 4$ µm at $T = 7$ K. We note that these values are lower than the ones reported for the electron-electron mean free path in [10], which was calculated with a much smaller grid resolution. However, they are comparable to the latter's estimate for the phonon-mediated interacting mean free path. While a full analysis is beyond the scope of this work, we point out that the combined momentum-conserving mean free path could thus be even slightly smaller. These estimated lower values of $l_{ee}$ give credence to the mechanism described in the main text whereby a weakly ballistic flow can effectively become hydrodynamic in thin samples.

**Transition from vortex to laminar flow in additional samples**

Analogously to the analysis in the main text (sample A), we show the transition from vortex to laminar flow in two additional samples, B and C. These samples provide additional insight into the dependence of the flow on the geometrical parameters, including the width the central strip, $W$, the chamber radius, $R$, and the sample thickness, $d$. Importantly, samples A and B were fabricated from the same batch of WTe$_2$ crystals, while sample C was exfoliated from a different batch of lower quality. AFM images of samples A, B, and C are shown in Extended Data Fig. 6.

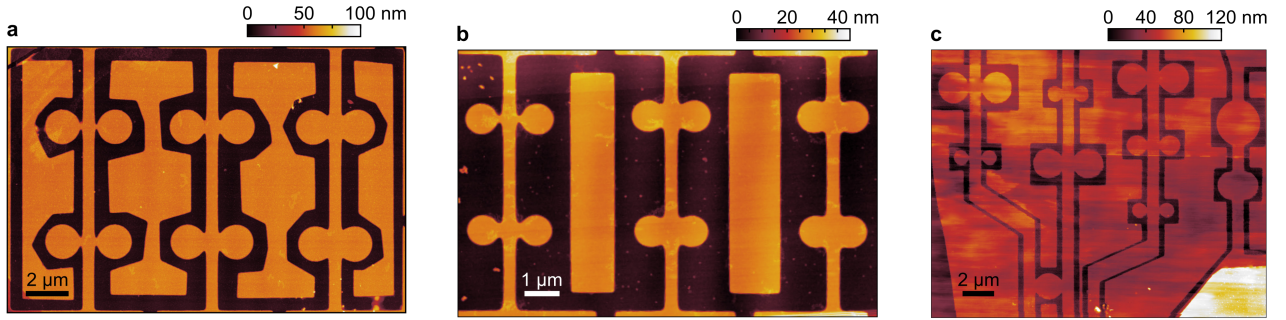

**Extended Data Fig. 6. AFM images of WTe$_2$ devices. a**, AFM image of device A analyzed in the main text with $W = 550$ nm, $R = 900$ nm, $d = 48$ nm, and aperture angles $\theta = 20°, 35°, 54°, 72°, 90°$, and $120°$. **b**, Device B used for Extended Data Fig. 7 with $W = 350$ nm, $R = 450$ nm, and $d = 23$ nm. **c**, Device C with $W = 770$ nm and $d = 30$ nm, and $R = 950, 725$, and $500$ nm (Extended Data Figs. 8) and dual-drive geometry at the bottom part (Extended Data Figs. 9).

Sample B is characterized by $W = 350$ nm, $R = 450$ nm, and $d = 23$ nm. To avoid current heating due to the narrower central strip, the excitation current was reduced to $I_0 = 25$ µA. Similar to the behavior in sample A in Fig. 4, the $J_x(x, y)$ images in Extended Data Figs. 9a,d,g show a transition from single-vortex to two-vortex to no-vortex state upon increasing $\theta$. This is in good agreement with simulations of $J_x(x, y)$ (Extended Data Figs. 7b,e,h) as well as simulated current streamlines (Extended Data Figs. 7c,f,i). Here, the two-vortex state occurs at $\theta_t = 60°$ as compared to $\theta_t = 54°$ in sample A. This aperture angle allows us to deduce $D/W = 0.35$ and resulting $D = 123$ nm in sample B. Thus, the smaller strip width $W$ of sample B results in an effective vertical upshift of the data points in Fig. 3a.

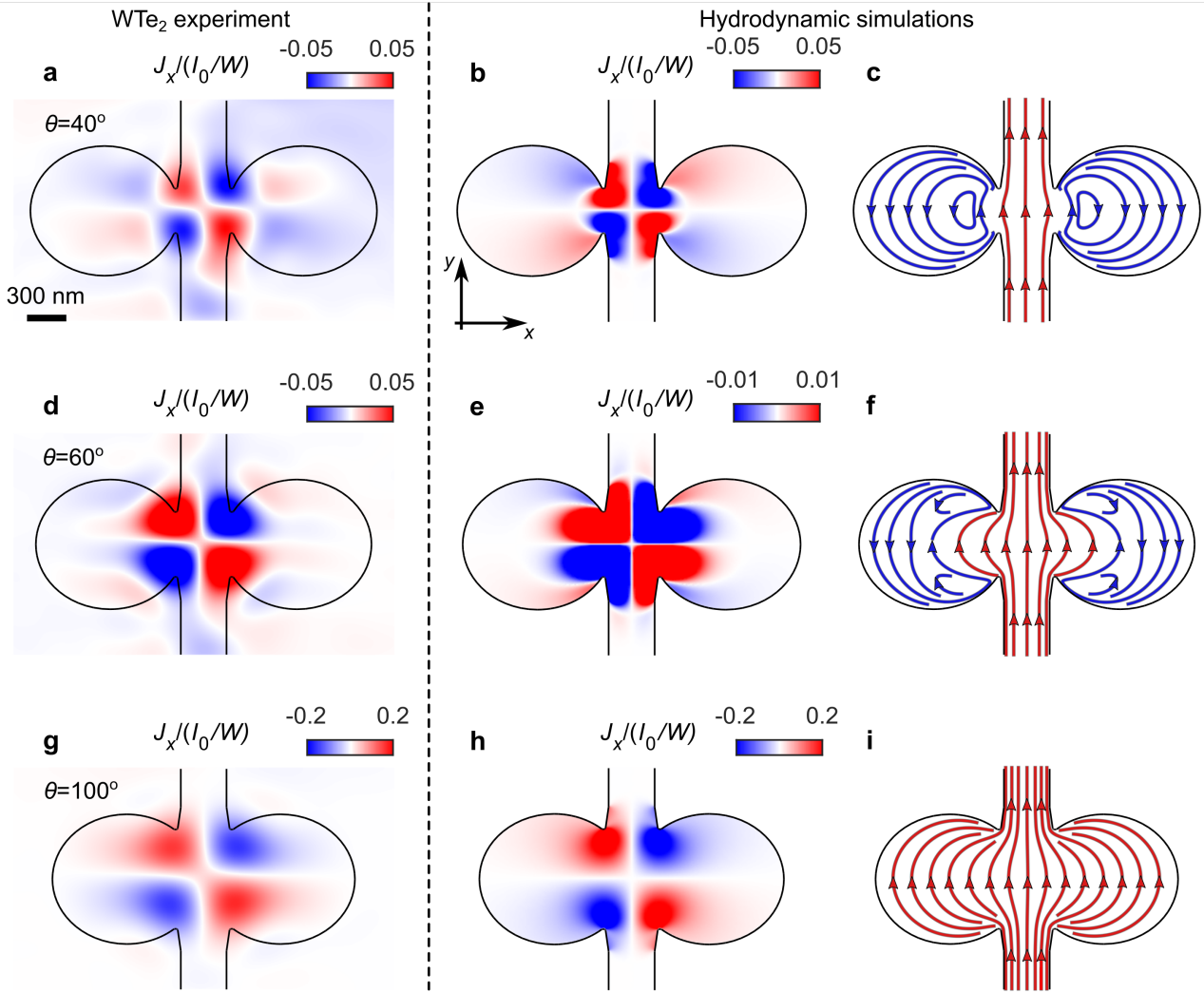

**Extended Data Fig. 7. Transition from single-vortex to two-vortex to laminar flow in sample B. a-c**, Measurement of single vortex state in device B with $\theta = 44°$ and corresponding simulations in the hydrodynamic regime with $D = 123$ nm and $\xi = 200$ nm. **a**, Measured current density $J_y(x, y)$ normalized by $I_0/W$ at $I_0 = 25$ μA. **b**, Simulated $J_y(x, y)$. **c**, Simulated current streamlines showing laminar (red) flow in the central strip and vortex flow (blue) in the chambers. **d-f**, Same as (a-c) for $\theta = 60°$ showing banana-shaped vortex at the transition from a single to double-vortex state. **g-i**, Same as (a-c) for $\theta = 100°$, showing laminar flow.

Extended Data Fig. 8 shows $J_x(x, y)$ distributions in sample C, which is characterized by $W = 770$ nm and $d = 30$ nm. Here, the disc chambers of radius $R = 950$ nm show single-vortex state for aperture angles of $\theta = 24°$ (a), $\theta = 45°$ (b), and $\theta = 60°$ (c), and laminar flow for $\theta = 180°$ (d). Although, no chamber with $\theta$ between 60° and 180° was available in this sample, the presence of a single vortex at $\theta = 60°$ indicates $D/W > 0.35$. The resulting $D > 270$ nm significantly exceeds the values derived for samples A and B. This may be attributed to the different batches of source material with different microscopic parameters or to variations in fabrication resulting in differences in the top and bottom surface quality.

In addition, single-vortex behavior was found in chambers with $R = 725$ nm and $\theta = 45°$ as well as for $R = 500$ nm and $\theta = 60°$ (Extended Data Figs. 8e,f), which shows the stability of the single-vortex state with respect to chamber radius.

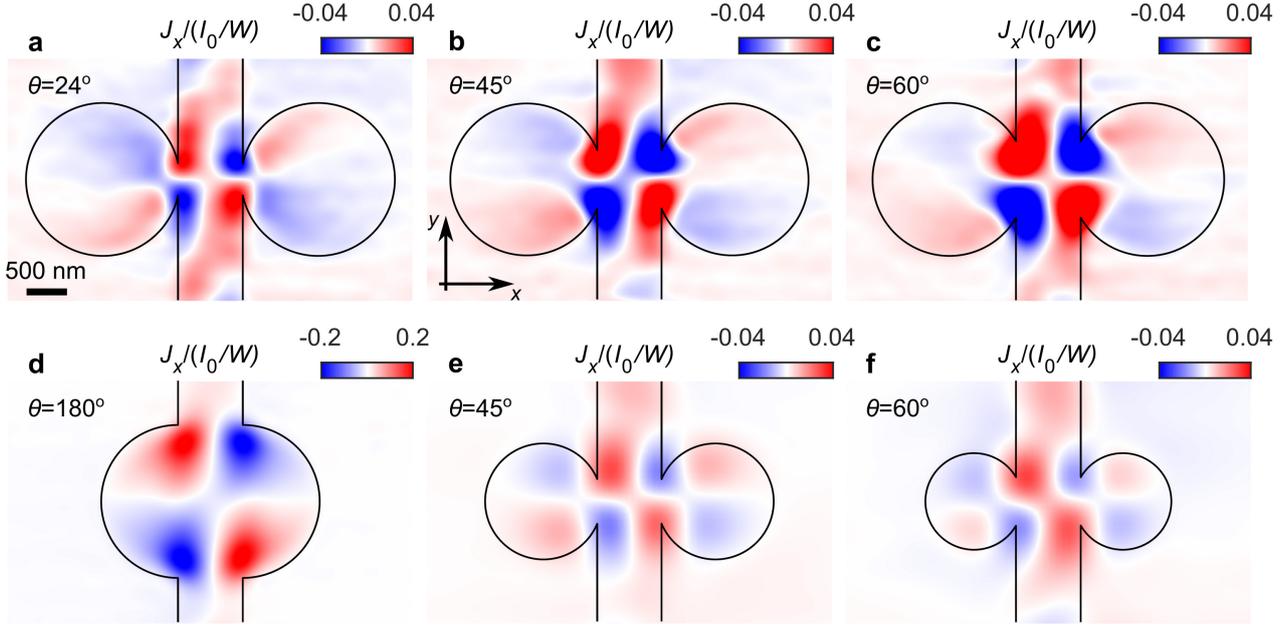

**Extended Data Fig. 8. Vortex flow in sample C with different geometrical parameters.** Current density $J_x(x,y)$ in sample C with $W = 770$ nm and $d = 30$ nm and various chamber parameters: $\theta = 24°$ and $R = 950$ nm (**a**), $\theta = 45°$ and $R = 950$ nm (**b**), $\theta = 60°$ and $R = 950$ nm (**c**), $\theta = 180°$ and $R = 950$ nm (**d**), $\theta = 45°$ and $R = 725$ nm (**e**) and $\theta = 60°$ and $R = 500$ nm (**f**). Laminar flow is observed in (d), while vortex flow is present in all the rest of the geometries.

### Dual-drive geometry

We have also studied an alternative geometry of a central disk with two apertures with $\theta = 44°$ on opposite sides connected to two current-driven strips patterned in WTe$_2$ sample C with $d = 30$ nm, and in Au film of similar thickness, as shown in Extended Data Fig. 9. An *ac* current of $I_L = 50$ µA was applied to the left strip with source at the bottom and drain at the top. A lower excitation frequency of $f = 17.73$ Hz was used to reduce capacitive currents between the two strips. A separate floating current source was used to apply current to the right strip with three values, $I_R = 0, -50$ µA, and 50 µA.

Panels a$_1$ and a$_2$ in Extended Data Fig. 9 show the $J_y(x,y)$ and $J_x(x,y)$ current distributions in Au sample for $I_R = 0$. In the ohmic regime, the current penetrates substantially into the central chamber similar to Fig. 1j. In the hydrodynamic case of WTe$_2$ (panel b$_1$), in contrast, the $J_y$ is mostly concentrated along the left strip with little protrusion into the chamber, analogous to Fig. 2g and consistent with the numerical simulations in panels g$_1$ and h$_1$. The transverse current, $J_x(x,y)$, reveals a laminar flow in Au (panel a$_2$) and a vortical flow in WTe$_2$ (panel b$_2$) in the chamber, consistent with the numerical simulations in panels g$_2$ and h$_2$, and with the simulated streamlines in panels g$_3$ and h$_3$. This configuration is equivalent to Figs. 1m and 2j of the main text.

Upon applying $I_R = -50$ µA, interesting flow patterns are observed in the central chamber. In the ohmic case, instead of flowing in and out of the chamber as observed for a single drive in panel a$_2$, part of the current traverses the chamber in its lower part (red $J_x$ in panel c$_2$), and then flows down to the bottom drain of the right strip as corroborated by the simulated streamlines in panel i$_3$. In the top half of the chamber, an opposite flow from top source of the right strip to the top drain on the left strip occurs (blue $J_x$), exchanging part of the currents from the two sources. Remarkably, in the hydrodynamic case, the shear forces of counter propagating currents in the two strips add up constructively and propel a single massive clockwise vortex in the entire chamber as observed in panel d$_2$ and simulated in panel j$_2$. The simulated streamlines in panel j$_3$ show that in this case the currents of the left and right sources do not mix: the laminar streams in the two strips are isolated by the whirlpool in the central chamber.

One would then expect that in the case of copropagating currents in the two strips (panel f₁), the opposing shear forces at the two apertures act destructively, annihilating the massive vortex. Instead, we find that a vortex–antivortex pair is nucleated in the chamber as visualized by $J_x(x,y)$ in panel f₂ and simulated in panels l₂ and l₃.

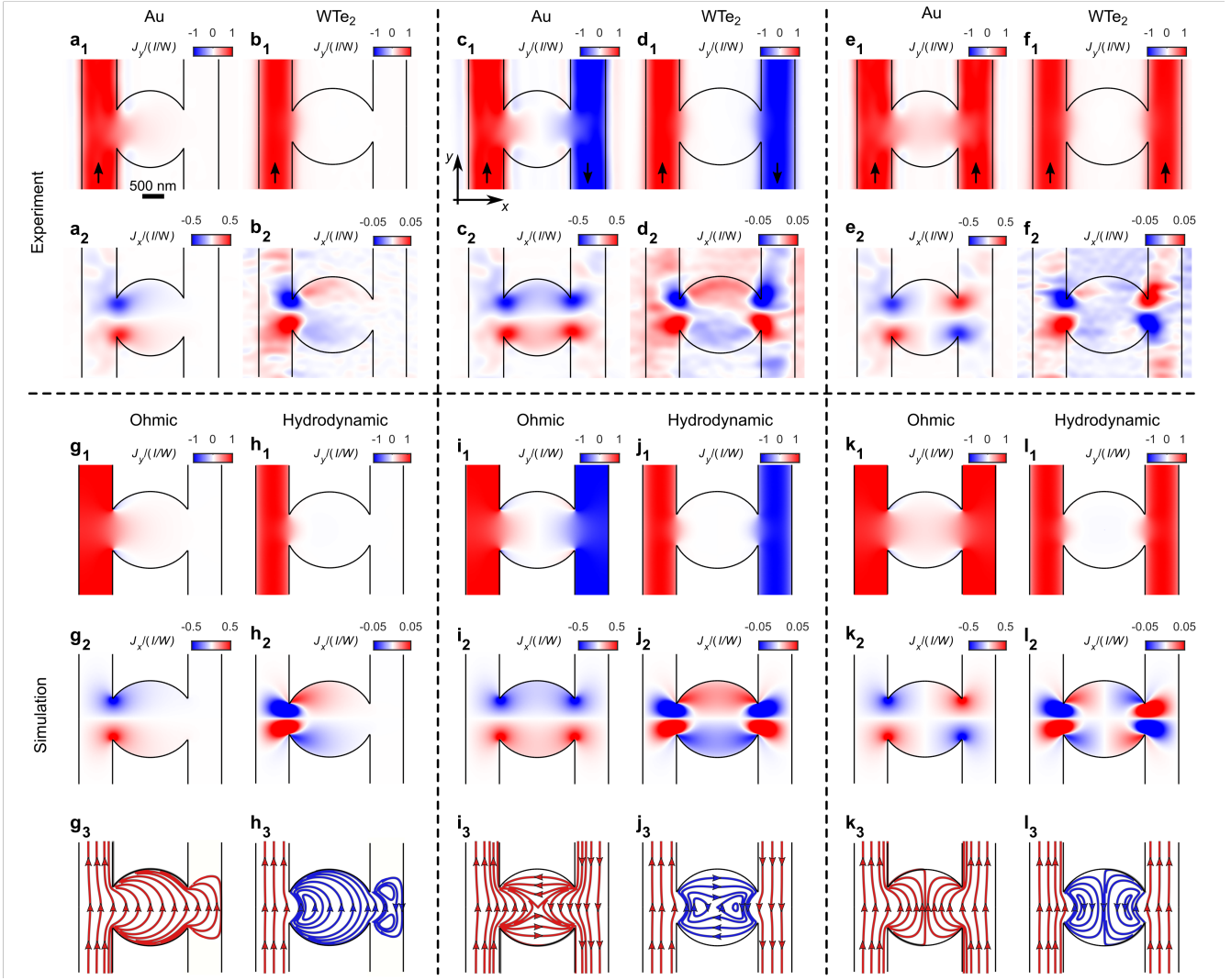

**Extended Data Fig. 9**. **Vortex–antivortex formation in dual-drive geometry. a-f**, Experimentally derived current densities $J_y(x,y)$ (top row) and $J_x(x,y)$ (bottom row) in Au and WTe₂ samples. **a-b**, Current $I_L = 50$ μA is driven in the up direction in the left strip with no current applied to the right strip resulting in a single vortex in the WTe₂ chamber in **b₂**. **c-d**, Counterpropagating currents $I_L = 50$ μA and $I_R = -50$ μA applied to the right and left strips, giving rise to a single massive vortex in **d₂**. **e-f**, Copropagating currents $I_L = 50$ μA and $I_R = 50$ μA applied to both strips which generates a vortex–antivortex pair in **f₂**. **g-l**, Numerical simulations of current densities $J_y(x,y)$ (top row), $J_x(x,y)$ (middle row) and the corresponding streamlines (bottom row) in the ohmic and hydrodynamic regimes for the three current configurations. The laminar streamlines are marked in red and the vortex streamlines in blue. The experimental data were acquired with pixel size of 10 nm, acquisition time of 40 ms/pixel, and image size of 600×350 pixels/image.

**Captions of Supplementary Videos**

**Supplementary Video 1 | Simulations of vortical-to-laminar flow transition in the para-hydrodynamic regime vs. $\theta$.** Numerical simulation of the current density $J_x(x,y)$ (top right) and the corresponding streamlines (bottom right) in the double-chamber geometry upon increasing the aperture angle $\theta$ for $D/W = 0.28$. The left panel shows the vortex stability phase diagram with no stress boundary conditions as presented in Fig. 3a. The purple dot marks the value of the varying $\theta$ along the $D/W = 0.28$ line. For $\theta \leq 54°$, there is a single vortex in each chamber (blue streamlines). Upon increasing $\theta$ further, the laminar flow (red streamlines) splits the single vortex in each chamber into two vortices, which are stable up to $\theta \leq 60°$. Finally, for $\theta > 60°$, the laminar streamlines fill the entire area of the chambers.

**Supplementary Video 2 | Simulations of vortical-to-laminar flow transition in the quasi-ballistic regime vs. $\theta$.** Numerical simulation of the current density $J_x(x,y)$ (top right) and the corresponding streamlines (bottom right) in the double-chamber geometry upon increasing $\theta$ for $D/W = 1.5$. The left panel shows the vortex stability phase diagram with no stress boundary conditions as presented in Fig. 3a. The purple dot marks the value of the varying $\theta$ along the $D/W = 1.5$ line. With increasing $\theta$, the laminar streamlines (red) gradually penetrate deeper into the chambers, distorting the vortices (blue streamlines) and pushing them towards the outer boundaries. The vortices become extinct at $\theta \cong 150°$ without splitting into double vortices as is the case in the hydrodynamic regime in Supplementary Video 1. For $\theta > 150°$, the laminar streamlines fill the entire area of the chambers.